\documentstyle[snow]{article}

\def\agt{
\mathrel{\raise.3ex\hbox{$>$}\mkern-14mu\lower0.6ex\hbox{$\sim$}}
}
\def\alt{
\mathrel{\raise.3ex\hbox{$<$}\mkern-14mu\lower0.6ex\hbox{$\sim$}}
}

\begin{document}
\bibliographystyle{unsrt}    

\newcommand{\st}{\scriptstyle}
\newcommand{\sst}{\scriptscriptstyle}
\newcommand{\mco}{\multicolumn}
\newcommand{\epp}{\epsilon^{\prime}}
\newcommand{\vep}{\varepsilon}
\newcommand{\ra}{\rightarrow}
\newcommand{\ppg}{\pi^+\pi^-\gamma}
\newcommand{\vp}{{\bf p}}
\newcommand{\ko}{K^0}
\newcommand{\kb}{\bar{K^0}}
\newcommand{\al}{\alpha}
\newcommand{\ab}{\bar{\alpha}}
\def\be{\begin{equation}}
\def\ee{\end{equation}}
\def\bea{\begin{eqnarray}}
\def\eea{\end{eqnarray}}
\def\CPbar{\hbox{{\rm CP}\hskip-1.80em{/}}}

\setcounter{secnumdepth}{2} 


\title{Gravitational Waves}

\firstauthors{ Kip S. Thorne}

\firstaddress{Theoretical Astrophysics, California Institute of
Technology, Pasadena, CA 91125, USA}

\secondauthors{}


\secondaddress{ $ $ }


\twocolumn[\maketitle
\null
\vskip-0.6truein
\abstracts{
This article reviews current efforts and plans for gravitational-wave
detection, the gravitational-wave sources that might be detected, and the
information that the detectors might extract from the observed waves.
Special attention is paid to (i) the LIGO/VIRGO network of earth-based,
kilometer-scale laser interferometers, which is now under construction
and will operate in the high-frequency band ($1$ to $10^4$ Hz),
and (ii) a proposed 5-million-kilometer-long Laser Interferometer Space
Antenna (LISA), which would fly in heliocentric orbit and
operate in the low-frequency band ($10^{-4}$ to $1$ Hz).  LISA would extend
the LIGO/VIRGO studies of stellar-mass ($M\sim2$ to $300 M_\odot$) black holes
into the domain of the massive black holes ($M\sim1000$ to $10^8M_\odot$)
that inhabit galactic nuclei and quasars.
}]

\section{Introduction}\label{introduction}

\footnote{This paper will be published in the
{\it Proceedings of the Snowmass
95 Summer Study on Particle and Nuclear Astrophysics and Cosmology},
eds. E. W. Kolb and R. Peccei (World Scientific, Singapore).}
According to general relativity theory, compact concentrations of energy
(e.g., neutron stars and black holes) should warp spacetime strongly, and
whenever such an energy concentration changes shape,
it should create a dynamically changing
spacetime warpage that propagates out
through the Universe at the speed of light.
This propagating warpage is called a {\it gravitational
wave}---a name that arises from general relativity's description of
gravity as a consequence of spacetime warpage.

Although gravitational waves have not yet been detected directly, their
indirect influence has been seen and measured with such remarkable accuracy
that their reality has been blessed even by the Nobel Prize Committee
(that bastion of conservatism which
explicitly denied Einstein the Prize for his relativity theories
\cite{pais}):

The 1993 Prize was awarded to
Russell Hulse and Joseph Taylor for their discovery of the binary pulsar
PSR 1913+16 \cite{hulse_taylor} and for Taylor's observational
demonstration (with
colleagues) \cite{taylor} that the binary's
two neutron stars are spiraling together at just the rate predicted
by general relativity's theory of gravitational radiation reaction:
from the observed orbit, one can compute the rate at which orbital energy
should be emitted into gravitational radiation, and from this rate of energy
loss one can compute the rate of orbital inspiral.  The computed and
observed inspiral rates agree to within the experimental accuracy,
better than one per cent.

Although this is a great triumph for Einstein, it
is not a firm proof that general relativity is correct in all respects.
Other relativistic theories of gravity (theories compatible with special
relativity) predict the existence of gravitational waves; and some other
theories predict the same inspiral rate for PSR 1913+16 as general relativity,
to within the experimental accuracy \cite{willbook,damour_taylor}.
Nevertheless, the experimental evidence for general relativity is so
strong \cite{willbook}, that I shall assume it to be correct
throughout this lecture except for occasional side remarks.

There are a number of efforts, worldwide, to detect gravitational
radiation.  These efforts are driven in part by the desire to
``see gravitational waves in the flesh,'' but more importantly by
the goal of using the
waves as a probe of the Universe and of the nature of gravity.  And a
powerful probe they should be, since they carry detailed
information about gravity and their sources.

There is an enormous difference between gravitational waves, and the
electromagnetic waves on which our present knowledge of the Universe is
based:
\begin{itemize}
\item
Electromagnetic waves are oscillations of the electromagnetic field that
propagate through spacetime; gravitational waves are oscillations of the
``fabric'' of spacetime itself.
\item
Astronomical electromagnetic waves are almost always incoherent
superpositions of
emission from individual electrons, atoms, or molecules.
Cosmic gravitational waves are produced
by coherent, bulk motions of huge amounts of mass-energy---either material
mass, or the energy of vibrating, nonlinear spacetime curvature.
\item
Since the wavelengths of electromagnetic waves are small compared to their
sources (gas clouds, stellar atmospheres, accretion disks,
...), from the waves we can make
pictures of the sources.  The wavelengths of cosmic gravitational waves
are comparable to or larger than their coherent, bulk-moving sources, so
we cannot make pictures from them. Instead, the gravitational waves
are like sound; they carry, in two independent waveforms,
a stereophonic, symphony-like description of their sources.
\item
Electromagnetic waves are easily absorbed, scattered, and dispersed by matter.
Gravitational waves travel nearly unscathed through all
forms and amounts of intervening matter \cite{300yrs,leshouches}.
\item
Astronomical electromagnetic waves have frequencies that begin at $f\sim
10^7$ Hz and extend on {\sl upward} by roughly 20 orders of magnitude.
Astronomical gravitational waves should begin at $\sim 10^4$ Hz
(1000-fold lower than the lowest-frequency astronomical electromagnetic waves),
and should
extend on {\sl downward} from there by roughly 20 orders of magnitude.
\end{itemize}

These enormous differences make it likely that:
\begin{itemize}
\item
The information brought to us by
gravitational waves will be very different from (almost ``orthogonal to'')
that carried by electromagnetic waves; gravitational waves will show us
details of the bulk motion of dense concentrations of energy, whereas
electromagnetic waves show us the thermodynamic state of optically thin
concentrations of matter.
\item
Most (but not all)
gravitational-wave sources that our instruments
detect will not be seen electromagnetically, and
conversely, most objects observed electromagnetically will never be seen
gravitationally.
Typical electromagnetic sources are
stellar atmospheres, accretion disks, and clouds of
interstellar gas---none of which emit significant gravitational waves,
while typical gravitational-wave sources are the cores of supernovae
(which are hidden from electromagnetic view by dense layers of
surrounding stellar gas), and colliding black holes (which emit no
electromagnetic waves at all).
\item
Gravitational waves may bring us great surprises.
In the past, when a radically new window has been opened onto the
Universe, the resulting surprises have had a profound, indeed
revolutionary, impact.  For example, the radio universe, as discovered in
the 1940s, 50s and 60s, turned out to be far more violent than the optical
universe; radio waves brought us quasars, pulsars, and the cosmic
microwave radiation, and with them our first direct observational
evidence for black holes, neutron stars, and the heat of the big bang
\cite{sullivan}.
It is reasonable to hope that gravitational waves will bring a similar
``revolution''.
\end{itemize}

In this lecture I shall review the present status of attempts to
detect gravitational radiation and plans for the future, and I shall
describe some examples of information that we expect to garner from
the observed waves.  I shall begin, in Section \ref{bands}, with an overview
of all the frequency bands in which astrophysical gravitational waves
are expected to be strong, the expected sources in each band, and the
detection techniques being used in each.  Then in subsequent sections I
shall focus on (i) the ``high frequency band'' which is populated by waves
from stellar mass black holes and neutron stars and is being probed by
ground-based instruments: laser interferometers and resonant-mass
antennas (Sections  \ref{gbint}, \ref{resonantbars},
\ref{coalescing_binaries}, and \ref{otherhfsources}),
and (ii) the ``low-frequency band'' which
is populated by waves from supermassive black holes and binary stars and
is probed by space-based instruments: radio and optical tracking of
spacecraft (Sections \ref{lisa} and \ref{lfsources}).  Finally, in
Section \ref{stochastic} I shall describe the stochastic background of
gravitational waves that is thought to have been produced by various
processes in the early universe, and prospects for detecting it in the
various frequency bands.

\section{Frequency Bands, Sources,
and Detection Methods}\label{bands}

Four gravitational-wave frequency bands are being explored
experimentally: the high-frequency band
(HF; $f\sim 10^4$ to $1$ Hz), the low-frequency band (LF; $f \sim 1$
to $10^{-4}$ Hz), the very-low frequency band (VLF; $f \sim 10^{-7}$ to
$10^{-9}$ Hz), and the extremely-low-frequency band (ELF; $f \sim
10^{-15}$ to $10^{-18}$ Hz).

\subsection{High-Frequency Band, 1 to $10^4$ Hz}\label{hf}

A gravitational-wave source of mass $M$ cannot be much smaller than its
gravitational radius, $2GM/c^2$, and cannot emit strongly at periods
much smaller than the light-travel time $4\pi GM/c^3$ around this gravitational
radius. Correspondingly, the frequencies at which it
emits are
\begin{equation}
f \alt {1\over{4\pi GM/c^3}} \sim 10^4 {\rm Hz}{M_\odot\over M}\;,
\label{fmax}
\end{equation}
where $M_\odot$ is the mass of the Sun and $G$ and $c$ are Newton's
gravitation constant and the speed of light.  To achieve a size of order
its gravitational radius and thereby emit near this maximum frequency,
an object presumably
must be heavier than the Chandrasekhar limit, about the mass of the sun,
$M_\odot$.
Thus, the highest frequency expected for strong gravitational
waves is $f_{\rm max} \sim 10^4$ Hz.
This defines the upper edge of the high-frequency gravitational-wave band.

The high-frequency band is the domain of Earth-based gravitational-wave
detectors: laser interferometers and resonant mass antennas.  At frequencies
below about 1 Hz, Earth-based detectors face nearly insurmountable noise
(i) from fluctuating Newtonian gravity gradients (due, e.g., to the
gravitational pulls of inhomogeneities in the Earth's atmosphere which
move overhead with the wind), and (ii) from Earth vibrations
(which are extremely difficult to filter out mechanically below $\sim 1$
Hz).  This defines the $1$ Hz lower edge of the high-frequency band; to
detect waves below this frequency, one must fly one's detectors in
space.

A number of interesting gravitational-wave sources fall in the
high-frequency band:  the stellar collapse to a neutron star or black
hole in our Galaxy and distant galaxies, which sometimes
triggers supernovae; the
rotation and vibration of neutron
stars (pulsars) in our Galaxy; the coalescence of neutron-star and
stellar-mass black-hole binaries ($M \alt 1000 M_\odot$)
in distant galaxies; and possibly such sources of
stochastic background as vibrating loops of cosmic string, phase
transitions in the early Universe, and the big bang in which the
Universe was born.

I shall discuss the high-frequency band in detail in Sections
\ref{gbint}--\ref{otherhfsources}.

\subsection{Low-Frequency Band, $10^{-4}$ to 1 Hz}\label{lf}

The low-frequency band, $10^{-4}$  to 1 Hz, is the domain of detectors
flown in space (in Earth orbit or in interplanetary orbit). The most
important of these are the Doppler tracking of spacecraft via
microwave signals sent
from Earth to the spacecraft and there transponded back to Earth (a
technique that NASA has pursued since the early 1970's), and optical
tracking of spacecraft by each other (laser interferometry in space, a
technique now under development for possible flight in $\sim 2014$ or
sooner).

The 1 Hz upper edge of the low-frequency band is defined by the
gravity-gradient and seismic cutoffs on Earth-based instruments; the
$\sim 10^{-4}$ Hz lower edge is defined by expected severe difficulties
at lower frequencies
in isolating spacecraft from the buffeting forces of fluctuating solar
radiation pressure, solar wind, and cosmic rays.

The low-frequency band should be populated by waves from short-period
binary stars in our own Galaxy (main-sequence binaries, cataclysmic
variables, white-dwarf binaries, neutron-star binaries, ...); from
white dwarfs, neutron stars, and small black holes spiraling into
massive black holes ($M \sim 3\times 10^5$ to $3\times
10^7 M_{\odot}$) in distant
galaxies; and from the inspiral and coalescence of supermassive
black-hole binaries ($M\sim100$ to $10^8 M_\odot$).
The upper limit, $\sim 10^8 M_\odot$, on the
masses of black holes that can emit in the low-frequency band is set by
Eq.\ (\ref{fmax}) with $f\agt 10^{-4}$ Hz.  There should also be a
low-frequency stochastic background from such early-universe processes
as vibrating cosmic strings, phase transitions, and the big-bang
itself.

I shall discuss the low-frequency band in detail in Sections \ref{lisa}
and \ref{lfsources}.

\subsection{Very-Low-Frequency Band, $10^{-7}$ to $10^{-9}$
Hz}\label{vlf}

Joseph Taylor and others have achieved a remarkable gravity-wave
sensitivity in the very-low-frequency band (VLF) by the timing of millisecond
pulsars: When a gravitational wave passes over the Earth, it perturbs
our rate of flow of time and thence the ticking rates of our clocks
relative to clocks outside the wave.  Such perturbations will show up
as apparent fluctuations in the times of arrival
of the pulsar's pulses.  If no fluctuations are seen at some
level, we can be rather sure that neither Earth nor the pulsar is being
bathed by gravitational waves of the corresponding strength.  If fluctuations
with the same time evolution
are seen simultaneously in the timing of several different
pulsars, then the cause could well be gravitational waves bathing the
Earth.

By averaging the pulses' times of arrival over long periods of time
(months to tens of years), a very high timing precision can be
achieved, and correspondingly tight limits can be placed on the waves
bathing the Earth
or the pulsar.  The upper edge of the VLF band, $\sim 10^{-7}$
Hz, is set by the averaging time, a few months, needed to build up high
accuracy; the lower edge, $\sim 10^{-9}$ Hz, is set by the time, $\sim 20$
years, since very steady millisecond pulsars were first discovered.

As we shall see in Section \ref{armlength},
strong gravitational-wave sources are generally
compact, not much larger than their own gravitational radii.
The only compact bodies that can radiate in the VLF band or below, i.e.,
at $f \alt 10^{-7}$ Hz, are those with $M \agt 10^{11} M_\odot$ [cf.\
Eq.\ (\ref{fmax})].
Conventional astronomical wisdom suggests that compact bodies this
massive do not exist, and that therefore the only strong waves in the
VLF band and below are a stochastic background produced by
the same early-universe processes as might radiate at low and high
frequencies: cosmic strings, phase transitions, and the big bang.

Of course, conventional wisdom could be wrong.  Nevertheless, it is
conventional to quote measurement accuracies in the VLF band and below
in the language of a stochastic background: the fraction $\Omega_g (f)$
of the energy required to close the universe that lies in a bandwidth
$\Delta f = f$ centered on frequency $f$.  The current $95\%$-confidence
limit on
$\Omega_g$ from pulsar timing in the VLF band is $\Omega_g(4\times10^{-9}{\rm
Hz}) < 6\times
10^{-8} H^{-2}$ where $H$ is the Hubble constant in units of 100 km
sec$^{-1}$ Mpc$^{-1}$ \cite{kaspi_taylor}.
This is a sufficiently tight limit that it is beginning to cast doubt
on the (not terribly popular) suggestion, that the Universe contains
enough vibrating loops of cosmic string for their gravitational pulls
to have seeded galaxy formation \cite{zeldovich_strings,vilenkin_strings}.

\subsection{Extremely-Low-Frequency Band, $10^{-15}$
to $10^{-18}$ Hz}
\label{elf}

Gravitational waves in the extremely-low-frequency band (ELF),
$10^{-15}$
to $10^{-18}$ Hz, should produce anisotropies in the cosmic
microwave background radiation.  The tightest limit from microwave
observations comes from the lower edge of the ELF band
$f\sim 10^{-18}$ Hz, where the gravitational wavelength is about
$\pi$ times the Hubble distance, and the waves, by squeezing all of the space
inside our cosmological horizon in one direction, and stretching it
all in another,
should produce a quadrupolar anisotropy in the microwave background.
The quadrupolar anisotropy measured by the COBE satellite, if due primarily
to gravitational waves (which it could be
\cite{krauss_white,davis}),
corresponds to an energy
density $\Omega_g (10^{-18}{\rm Hz}) \sim 10^{-9}$.
In Section \ref{stochastic} I shall discuss the
implications of this impressive ELF limit for the strength of the
early-Universe stochastic background in the HF and LF bands.

\subsection{Other Frequency Bands and Other Detection Methods}
\label{otherbands}

A large number of other methods have been conceived of, for searching
for gravitational radiation.  Some of them would operate best in the
HF, LF, VLF, and ELF bands described above; others would operate best
at other frequencies.  However, none has shown anywhere near the
promise or the achievements of the methods described above (laser
interferometry on Earth and in space, resonant mass antennas,
Doppler tracking of
spacecraft, timing of pulsars, and anisotropy of microwave background).
For some references to other methods, see, e.g., \cite{300yrs}.

\begin{figure}
\vskip 6.0pc
\special{hscale=45 vscale=45 hoffset=-2 voffset=-13
psfile=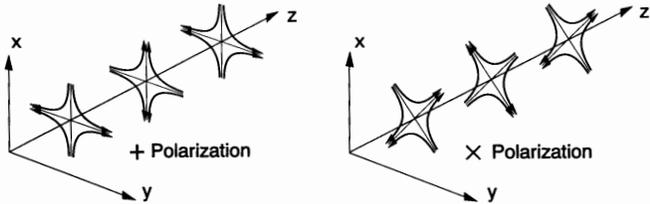}
\caption{The lines of force associated with the two polarizations of
a gravitational wave.  (From Ref. \protect\cite{ligoscience}.)
}
\label{fig:forcelines}
\end{figure}

\section{Ground-Based Laser Interferometers}\label{gbint}

\subsection{Wave Polarizations, Waveforms, and How an Interferometer
Works}
\label{intworks}

According to general relativity theory (which I shall assume to be
correct in this paper), a gravitational wave has two
linear polarizations,
conventionally called $+$ (plus) and $\times$ (cross).  Associated with
each polarization there is a gravitational-wave field, $h_+$ or $h_\times$,
which oscillates in time and propagates with the speed of light.  Each
wave field produces tidal forces (stretching and squeezing forces) on
any object or detector through which it passes.  If the object is small
compared to the waves' wavelength (as is the case for ground-based
interferometers and resonant mass antennas), then relative to the
object's center, the forces have the quadrupolar patterns shown in
Figure~\ref{fig:forcelines}.
The names ``plus'' and ``cross'' are derived from the orientations of the
axes that characterize the force patterns \cite{300yrs}.

\begin{figure}
\center
\vskip 9.2pc
\special{hscale=50 vscale=50 hoffset=-3 voffset=-5
psfile=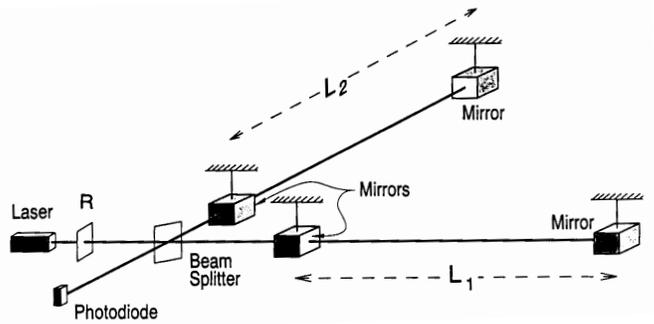}
\caption{Schematic diagram of a laser interferometer
gravitational wave detector.  (From Ref.\ \protect\cite{ligoscience}.)
}
\label{fig:interferometer}
\end{figure}

A laser interferometer gravitational wave detector (``interferometer'' for
short) consists of four masses that hang from vibration-isolated
supports as shown in Figure \ref{fig:interferometer}, and the indicated
optical system for
monitoring the separations between
the masses \cite{300yrs,ligoscience}. Two masses are near
each other, at the corner of an ``L'',
and one mass is at the end of each of the L's long arms.  The arm
lengths are nearly equal, $L_1 \simeq L_2 = L$.  When a gravitational
wave, with frequencies high compared to the masses' $\sim 1$ Hz pendulum
frequency, passes through the detector, it pushes the masses back and
forth relative to each other as though they were free from their
suspension wires, thereby changing the arm-length difference,
$\Delta L \equiv L_1-L_2$.  That change is
monitored by laser interferometry in such a way that the variations in
the output of the
photodiode (the interferometer's output) are directly proportional to
$\Delta L(t)$.

If the waves are coming from overhead or
underfoot
and the axes of the $+$ polarization coincide with the arms'
directions, then it is the waves' $+$ polarization that drives the masses, and
$\Delta L(t) / L = h_+(t)$.  More generally,
the interferometer's output is a linear combination of the two
wave fields:
\begin{equation}
{\Delta L(t)\over L} = F_+h_+(t) + F_\times h_\times (t) \equiv h(t)\;.
\label{dll}
\end{equation}
The coefficients $F_+$ and $F_\times$ are of order unity and depend in a
quadrupolar manner on
the direction to the source and the orientation of the detector \cite{300yrs}.
The combination $h(t)$ of the two $h$'s is called the
{\it gravitational-wave strain} that acts on the detector; and the time
evolutions of $h(t)$, $h_+(t)$, and $h_\times(t)$ are sometimes called
{\it waveforms}.

Interferometer test masses at present are made of
transparent fused silica, though other materials might be used
in the future.  The masses' inner faces (shown white
in Fig.\ \ref{fig:interferometer}) are
covered with high-reflectivity dielectric coatings
to form the indicated ``mirrors'',
while the masses' outer faces are covered with anti-reflection coatings.
The two mirrors facing each other on each arm form a Fabry-Perot
cavity.  A beam splitter splits a carefully prepared laser beam in two,
and directs the resulting beams down the two arms. Each beam penetrates
through the antireflection coating of its arm's corner mass,
through the mass, and through the dielectric coating (the mirror);
and thereby---with the length of the arm's Fabry-Perot cavity adjusted
to be nearly an integral number of half wavelengths of light---the beam gets
trapped in the cavity.
The cavity's end mirror has much higher reflectivity than
its corner mirror, so the trapped light leaks back out through the
corner mirror, and then hits the beam splitter where it recombines with
light from the other arm.  Most of the recombined light goes back toward
the laser (where it can be returned to the interferometer by a
``light-recycling mirror'' labeled $R$), but a tiny portion
goes toward the photodiode.

When a
gravitational wave hits the detector and moves the masses, thereby
changing the lengths $L_1$ and $L_2$ of the two cavities, it shifts each
cavity's resonant frequency slightly relative to the laser frequency, and
thereby changes the phase of the light in the cavity and the phase of
the light that exits from the cavity toward the beam splitter.
Correspondingly, the relative phase of the two beams returning to the
splitter is altered by an amount $\Delta\Phi \propto \Delta L$, and this
relative phase shift causes a change in the intensity of the recombined
light at the photodiode, $\Delta I_{\rm pd} \propto \Delta\Phi \propto \Delta L
\propto h(t)$.  Thus, the change of photodiode output current is directly
proportional to the gravitational-wave strain $h(t)$.
This method of monitoring $h(t)$, which
was invented by Ronald Drever \cite{drever} as a modification of
Rainer Weiss's \cite{weiss}
seminal concept for such an interferometer, is capable of very high
sensitivity, as we shall see below.

\subsection{Wave Strengths and Interferometer Arm Lengths}
\label{armlength}

The strengths of the waves from a gravitational-wave source can be
estimated using the
``Newtonian/quadrupole'' approximation to the Einstein field equations.
This approximation says that $h\simeq (G/c^4)\ddot Q/r$, where
$\ddot Q$ is the second time derivative of the source's quadrupole
moment, $r$ is the distance of the source from Earth (and $G$ and $c$
are Newton's gravitation constant and the speed of light).
The strongest sources will be highly nonspherical and thus will
have $Q\simeq ML^2$, where $M$ is their mass and $L$ their size, and
correspondingly will have $\ddot Q \simeq 2Mv^2 \simeq 4 E_{\rm
kin}^{\rm ns}$, where $v$ is their internal velocity and
$E_{\rm kin}^{\rm ns}$ is the nonspherical part of their
internal kinetic energy.  This provides us with the estimate
\begin{equation}
h\sim {1\over c^2}{4G(E_{\rm kin}^{\rm ns}/c^2) \over r}\;;
\label{hom}
\end{equation}
i.e., $h$ is about 4 times the gravitational potential produced at Earth by
the mass-equivalent of the source's nonspherical, internal kinetic
energy---made dimensionless by dividing by $c^2$.  Thus, in order to
radiate strongly, the source must have a very large, nonspherical,
internal kinetic energy.

The best known way to achieve a huge internal kinetic energy is via
gravity; and by energy conservation (or the virial theorem), any
gravitationally-induced kinetic energy must be of order the source's
gravitational potential energy.  A huge potential energy, in turn,
requires that the source be very compact, not much larger than its own
gravitational radius.  Thus, the strongest gravity-wave sources must be
highly compact, dynamical concentrations of large amounts of mass (e.g.,
colliding and coalescing black holes and neutron stars).

Such sources cannot remain highly dynamical for long; their motions will
be stopped by energy loss to gravitational waves and/or the formation of
an all-encompassing black hole.  Thus, the strongest sources should be
transient.  Moreover, they should be very rare --- so rare that to see a
reasonable event rate will require reaching out through a substantial
fraction of the Universe.  Thus, just as the strongest radio waves
arriving at Earth
tend to be extragalactic, so also the strongest gravitational waves are
likely to be extragalactic.

For highly compact, dynamical objects that radiate in the high-frequency band,
e.g.\ colliding and coalescing neutron stars and stellar-mass black
holes, the internal, nonspherical kinetic energy
$E_{\rm kin}^{\rm ns}/c^2$ is of order the mass of
the Sun; and, correspondingly, Eq.\ (\ref{hom}) gives $h\sim 10^{-22}$
for such sources at the Hubble
distance (3000 Mpc, i.e., $10^{10}$ light years);
$h\sim 10^{-21}$ at 200 Mpc (a best-guess distance for several
neutron-star coalescences per year; see Section \ref{coalescencerates}),
$h\sim 10^{-20}$ at the
Virgo cluster of galaxies (15 Mpc); and $h\sim 10^{-17}$ in the outer
reaches of our own Milky Way galaxy (20 kpc).  These numbers set the
scale of sensitivities that ground-based interferometers seek to achieve:
$h \sim 10^{-21}$ to $10^{-22}$.

When one examines the technology of laser interferometry, one sees good
prospects to achieve measurement accuracies $\Delta L \sim 10^{-16}$ cm
(1/1000 the diameter of the nucleus of an atom).  With such an accuracy,
an interferometer must have an arm length $L = \Delta L/h \sim 1$ to 10
km, in order to achieve the desired wave sensitivities, $10^{-21}$ to
$10^{-22}$.  This sets the scale of the interferometers that are now
under construction.

\subsection{LIGO, VIRGO, and the International Interferometric Network}
\label{network}

\begin{figure}
\vskip 11.6pc
\special{hscale=51 vscale=51 hoffset=0 voffset=-26
psfile=artistsligo.eps}
\caption{Artist's conception of one of the LIGO interferometers.
[Courtesy the LIGO Project.]}
\label{fig:LIGOart}
\end{figure}

Interferometers are plagued by non-Gaussian noise, e.g.\ due
to sudden strain releases in the wires that suspend the masses.  This
noise prevents a single interferometer, by itself, from detecting with
confidence short-duration gravitational-wave bursts (though it might
be possible for a single interferometer to search for the periodic
waves from known pulsars).   The non-Gaussian noise can be removed
by cross correlating two, or preferably three or more, interferometers
that are networked together at widely separated sites.

The technology and techniques for such interferometers have been under
development for nearly 25 years, and plans for km-scale
interferometers have been developed over the past 14 years.  An
international network consisting of three km-scale interferometers, at
three widely separated sites, is now in the early stages of
construction. It includes two sites of
the American LIGO Project (``Laser Interferometer Gravitational Wave
Observatory'') \cite{ligoscience}, and one site of the French/Italian VIRGO
Project (named after the Virgo cluster of galaxies) \cite{virgo}.

LIGO will consist of two
vacuum facilities with 4-kilometer-long arms, one in Hanford, Washington
(in the northwestern United States; Fig.\ \ref{fig:LIGOart})
and the other in Livingston, Louisiana (in the southeastern United States).
These facilities are designed to house
many successive generations of interferometers without the necessity of
any major facilities upgrade; and after a planned
future expansion,
they will be able to house several interferometers at once, each
with a different optical configuration optimized for a different type of
wave (e.g., broad-band burst, or narrow-band periodic wave, or
stochastic wave).  The LIGO facilities and their first interferometers
are being constructed by a team of about 80 physicists and engineers at
Caltech and MIT, led by Barry Barish (the PI) and Gary Sanders (the
Project Manager).  Robbie Vogt (who directed the project
during the pre-construction phase) is in charge of the final design and
construction of LIGO's first interferometers, Stan Whitcomb is in charge
of interferometer R\&D, and Albert Lazzarini is the system engineer and
Rai Weiss the cognizant scientist for integration of all parts of LIGO.

A number of other research groups
are making important contributions to LIGO:  Bob Byers' group at
Stanford is developing Nd:YAG lasers, Peter Saulson's group at Syracuse
and Vladimir Braginsky's group in Moscow are developing test-mass suspension
systems and studying noise in them; Jim Faller's group at JILA is
developing active vibration isolation systems; Ron Drever's group at
Caltech is developing advanced interferometers; and Sam Finn's group at
Northwestern and my group at Caltech are developing data analysis
techniques.  A number of other groups are likely to join the LIGO effort
in the next few years. A formal association of LIGO-related
scientists (the {\sl LIGO Research Community}, an analog of a
``user's group'' in high-energy physics) is
being organized, and a LIGO Program Advisory Committee will be formed
soon, with voting membership restricted to people outside the Caltech/MIT
LIGO team, to advise the LIGO management.

The VIRGO Project is building one vacuum facility in Pisa, Italy, with
3-kilometer-long arms.  This facility and its first interferometers are
a collaboration of more than a hundred physicists and engineers
at the INFN (Frascati, Napoli, Perugia, Pisa), LAL (Orsay), LAPP
(Annecy), LOA (Palaiseau), IPN (Lyon), ESPCI (Paris), and the University
of Illinois (Urbana), under the leadership of Alain Brillet and
Adalberto Giazotto.

Both LIGO and VIRGO are scheduled for completion
in the late 1990s, and their first gravitational-wave searches are
likely to be performed in 2000 or 2001.

LIGO alone, with its two sites which have parallel arms, will be able to
detect an incoming gravitational wave, measure one of its two waveforms,
and (from the time delay between the two sites) locate its source to within a
$\sim 1^{\rm o}$ wide annulus on the sky.
LIGO and VIRGO together, operating as a {\sl coordinated international
network}, will be able to locate the source
(via time delays plus the interferometers' beam patterns)
to within a 2-dimensional error box with size
between several tens of arcminutes and several degrees, depending on
the source direction and on
the amount of high-frequency structure in the waveforms.
They will also be able to monitor both waveforms $h_+(t)$ and
$h_\times(t)$ (except for frequency components above about 1kHz
and below about 10 Hz, where the interferometers' noise becomes severe).

The accuracies of the direction measurements and the ability to
monitor more than one
waveform will be severely compromised when the source lies anywhere near
the plane formed by the three LIGO/VIRGO interferometer locations.  To
get good all-sky coverage will require a fourth interferometer at a site
far out of that plane; Japan and Australia would be excellent locations,
and research groups there are carrying out research and development on
interferometric detectors, aimed at such a possibility.  A 300-meter
prototype interferometer called TAMA is under construction in Tokyo,
and a 400-meter prototype called AIGO400 has been proposed for
construction north of Perth.

Two other groups are major players in this field, one in Britain led
by James Hough, the other in Germany, led by Karsten Danzmann. These
groups each
have two decades of experience with
prototype interferometers (comparable experience to the LIGO team
and far more than anyone else) and great expertise.  Frustrated by
inadequate financing for
a kilometer-scale interferometer, they
are constructing, instead, a 600 meter
system called GEO600 near Hannover, Germany.  Their goal is to develop,
from the outset, an interferometer with the sort of advanced design that
LIGO and VIRGO will attempt only as a ``second-generation''
instrument, and thereby achieve sufficient sensitivity
to be full partners in the international
network's first gravitational-wave searches; they then would offer a
variant of their
interferometer as a candidate for second-generation operation in the
much longer arms of LIGO
and/or VIRGO.  It is a seemingly audacious plan, but with their
extensive experience and expertise, the British/German collaboration
might pull it off successfully.

\subsection{Interferometer Development and Noise Sources}
\label{intdevelop}

It is not possible, in the short vacuum systems now available (arm
lengths $\le 40$ meters), to develop and test a multikilometer
interferometer as a single unit.  This is because the various noise
sources that plague an interferometer scale differently from each other
with length $L$ and with gravity-wave frequency $f$.  As a result, the
various components of the multikilometer interferometers, and the various
techniques to be used in them, are being developed and tested
separately in a number of different laboratories, and will only be
combined together into a single interferometer when the LIGO/VIRGO
vacuum systems are completed.

The best known of the LIGO-Project laboratories in which components and
techniques are being developed is
the 40-meter prototype interferometer at Caltech (Fig.\
\ref{fig:mark2}).  This prototype focuses on the development of methods and
components
to control ``displacement noise,'' i.e., those noise sources that push
the mirrored test masses back and forth as would a gravity wave.
The principal sources of displacement noise are {\it seismic vibrations} of
the ground beneath the interferometer (which are filtered out by the
masses' suspension wires and by
``isolation stacks'' made of successive layers of steel and rubber), and
{\it thermally-induced vibrations} of the test masses and of
the wires that suspend them (vibrations that are controlled by designing the
test masses and suspensions with great care and constructing them
from low-loss, i.e.\ high ``Q'', materials).

\begin{figure}
\vskip13.5pc
\special{hscale=50 vscale=50 hoffset=0 voffset=-36
psfile=40mphoto.eps}
\caption{The LIGO Project's 40-meter ``Mark II'' prototype interferometer
at Caltech.  This prototype went into operation in 1993.  It has much
larger vacuum chambers, to accommodate bigger and better seismic
isolation stacks, than those of the previous ``Mark I'' prototype
(which operated from the early 1980s to 1992).  [Courtesy the LIGO
Project.]
}
\label{fig:mark2}
\end{figure}

Among the LIGO Project's other laboratories, there is a shorter-armed
prototype-interferometer facility at MIT, which is devoted to developing
methods and
components for controlling noise in the phase of the interferometer's light
beams.  Since the gravity wave makes itself known by the phase shift
that it puts on the light of one interferometer arm relative to the
other, this phase noise can simulate a gravity wave.  Among the various
causes of phase noise, the one that is the most fundamental
is {\it photon shot noise} due to the random
times at which the light's photons arrive at the photodiode (cf.\ Fig.\
\ref{fig:interferometer}).

\begin{figure}
\vskip16.8pc
\special{hscale=35 vscale=35 hoffset=25 voffset=-10
psfile=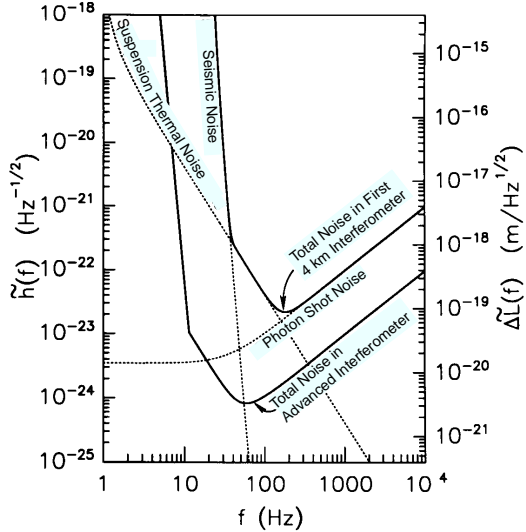}
\caption{The expected noise spectrum in each of LIGO's first 4-km
interferometers (upper solid curve) and in more advanced
interferometers (lower solid curve).  The dashed curves show various
contributions to the first interferometers' noise.  Plotted horizontally
is gravity wave frequency $f$; plotted vertically is $\tilde h(f)$,
the square root of
the spectral density of the detector's output $h(t) = \Delta L(t)/L$ in
the absence of a gravity wave.  The rms $h$ noise in a bandwidth $\Delta f$
at frequency $f$ is $h_{\rm rms} = \tilde h(f) \protect\sqrt{\Delta f}$.
(From Ref.\ \protect\cite{ligoscience}.)}
\label{fig:noisesources}
\end{figure}

Once the myriad of other noise sources have been brought under control,
{\it shot noise}, {\it thermal noise} (i.e., thermally induced
vibrations), and {\it seismic noise} (i.e., ground vibrations) are likely
to be the ultimate
impediments to detecting and studying gravitational waves.  Figure
\ref{fig:noisesources}
shows the spectra expected for each of these three noises
in the first interferometers that will operate in LIGO.  At frequencies
above 200 Hz, shot noise dominates; between 200 Hz and 40 Hz, thermal
noise in the suspension wires dominates; and below 40 Hz, seismic noise
dominates.

During LIGO's operations, step-by-step improvements will be made
in the control of these three noise sources \cite {ligoscience},
thereby pushing the overall
noise spectrum downward from the ``first-interferometer'' level
toward the ``advanced-interferometer''
level shown in Figure \ref{fig:noisesources}.  As we shall see below,
the sensitivity of the first interferometers
might be inadequate to detect gravitational waves.
However, we are quite confident that at some point
during the improvement
from first interferometers to advanced, a plethora of
gravitational
waves will be found and will start bringing us exciting information
about fundamental physics and the Universe.

Notice from Figure \ref{fig:noisesources} that the advanced LIGO
interferometers are expected to have
their optimal sensitivity at $f\sim 100$ Hz, and rather good sensitivity
all the way from $f\sim 10$ Hz at the low-frequency end to $f\sim 500$
Hz at the high-frequency end.  Below 10 Hz, seismic noise, creeping
through the isolation stacks, will overwhelm all gravitational-wave
signals; and above 500 Hz, photon shot noise may overwhelm the
signals.

\begin{figure}
\vskip14.4pc
\special{hscale=47 vscale=47 hoffset=-2 voffset=-5
psfile=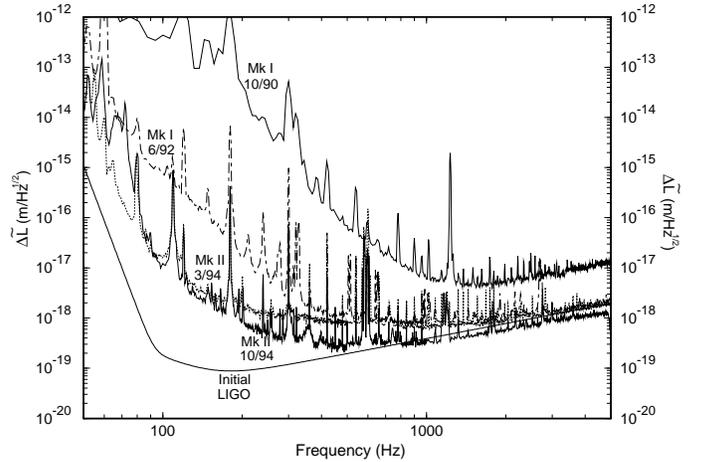}
\caption{Measured noise spectra in the Caltech 40-meter prototype
interferometer (Fig.\ \protect\ref{fig:mark2}).  Since this prototype is
devoted to learning to control displacement noise, the spectra shown are
$\Delta \tilde L(f)$, the square root of the spectral density of the
measured arm-length difference.  Each of the many spectral lines is
well understood, and most could be removed
if their removal were of high priority (e.g., they are multiples of the
60 Hz line frequency sneaking into the apparatus by known routes). Those
few, very narrow lines
that cannot be physically removed by cleaning up the instrument
(e.g., thermally-driven violin-mode resonances of the
wires that suspend the test masses) will be filtered out in the data
analysis.  Thus, the interferometer sensitivity is characterized by the
continuum noise floor and not the lines.
(From Ref.\ \protect\cite{40mnoisereport}.)}
\label{fig:40mspectrum}
\end{figure}

Figure \ref{fig:40mspectrum} gives an
impression of the present state of interferometer
technology and the rate of progress.  This figure shows a sequence of
noise spectra in the 40-meter prototype interferometer at Caltech,
during 1990--94 when the 40-meter R\&D emphasis was on improving the
low-frequency noise performance.
The top two (noisiest) spectra are snapshots of the original ``Mark I''
prototype performance in October 1990 and June 1992; the lower two
(quieter) spectra are from the rebuilt, ``Mark II'' prototype of Figure
\ref{fig:mark2}, in March and October 1994.  The smooth, solid line,
for comparison, is the
displacement noise goal for the first 4-km interferometers in LIGO
(i.e., it is the upper solid curve of Fig.\ \ref{fig:noisesources}).

Note that the prototypes's arm-length difference was being
monitored, in October 1994, to within an rms noise level (in a bandwidth
equal to frequency) $\Delta L_{\rm rms} = \sqrt f \Delta \tilde L(f) <
8\times 10^{-16}$ cm over the frequency range 200 to 1000 Hz.  This
corresponds to an rms gravitational-wave noise level $h_{\rm rms} =
\Delta L_{\rm rms}/40{\rm m} < 2\times 10^{-19}$, the best that any
gravitational-wave detector has yet achieved.

\subsection{Semiquantitative Discussion of Interferometer Noise}
\label{intnoise}

The LIGO and VIRGO interferometers are expected to have rms noise levels
$h_{\rm rms} \alt 10^{-22}$ corresponding to test-mass position noises
$\Delta L_{\rm rms} \alt hL \sim 10^{-16}$ cm.
$10^{-16}$ cm is awfully small:
$1/1000$ the diameter of the nucleus of an atom, and $10^{-12}$ the
wavelength of the light being used to monitor the masses' motions.
How can one possibly monitor such small motions?  The following estimate
explains how.

One adjusts the reflectivities of the interferometer's
corner mirrors so the two arms store the laser light on average
for about half a cycle of a $\sim 100$ Hz gravitational wave, which
means for $\sim 100$ round trips.  The light in each arm thereby
acquires a phase shift
\begin{equation}
\Delta \Phi \sim 100 \times 4\pi\Delta L/\lambda \sim 10^{-9}\;,
\label{dphi}
\end{equation}
where $\lambda \sim 10^{-4}$ cm is the wavelength of light.  If the
interference of the light from the two beams is done optimally, then
this phase shift (equal and opposite in the two arms) can be measured
at the photodiode
to an accuracy that is governed by the light's photon shot noise,
$\Delta\Phi \sim 1/\sqrt N$, where $N$ is the number of photons that enter the
interferometer from the laser during the half-cycle of photon storage
time.  (This $1/\sqrt N$ is the usual photon fluctuation in a
quantum mechanical ``coherent state'' of light.)  Thus, to achieve the
required accuracy, $\Delta \Phi \sim 10^{-9}$, in the face of photon
shot noise, requires $N\sim 10^{18}$ photons in 0.01 second, which means a
laser power of $\sim 100$ Watts.

By cleverness \cite{drever}, one can reduce the required laser power:  The
light is stored in the interferometer arms
for only a half gravity-wave period
($\sim 100$ round trips) because during the next
half period the waves would reverse the sign of $\Delta L$, thereby
reversing the sign of the phase shift being put onto the light and
removing from the light the signal that had accumulated
in the first half period.  In just 100 round trips, however, the light
is attenuated hardly at all.  One therefore reuses the light, over
and over again.  This is done by (i) operating the interferometer with only
a tiny fraction of the recombined light going out toward the photodiode,
and almost all of it instead going back toward the laser, and by
(ii) placing a mirror (marked $R$ in Fig.\ \ref{fig:interferometer})
between the laser and
the interferometer in just such a position that the entire
interferometer becomes an optical cavity driven by the laser---with its
arms as two subcavities.   Then the mirror $R$ recycles the recombined
light back into the interferometer in phase with the new
laser light,
thereby enabling a laser of, say, 5 Watts to behave like one of 100
Watts or more.

Turn from photon shot noise to thermal noise.  How, one might ask, can
one possibly expect to monitor the mirrors' motions at a level of
$10^{-16}$ cm when the room-temperature atoms of which the fused-silica
mirrors are made vibrate thermally with amplitudes $\Delta l_{\rm rms} =
\sqrt{kT/m\omega^2} \sim 10^{-10}$ cm?  (Here $k$ is
Boltzmann's constant, $T$ is room temperature, $m$ is the
atomic rest mass, and $\omega \sim 10^{14} \;{\rm s}^{-1}$ is the angular
frequency of atomic vibration.)  The answer is that these individual
atomic vibrations are
unimportant.  The light beam, with its $\sim 5$cm diameter,
averages over the positions of $\sim 10^{17}$ atoms in the mirror, and
with its $0.01$s storage time it averages over $\sim 10^{11}$
vibrations of each atom.  This spatial and temporal
averaging makes the vibrations of
individual atoms irrelevant.  Not so irrelevant, however, are the
lowest-frequency normal-mode vibrations
of the mirror-endowed masses (since these modes experience much less time
averaging than the faster atomic vibrations).  Assuming a mass
$m\sim ($a few tens of kg), these normal modes
have angular frequencies $\omega \sim 10^5 {\rm s}^{-1}$, so their
rms vibration amplitude is $\Delta l_{\rm rms}
= \sqrt{kT/m\omega^2}\sim 10^{-14}$ cm.  This
is 100 times larger than the signals we wish to monitor; but if these
modes have high quality factors (high $Q$'s; low losses), then the
vibrations will be very steady over the interferometer's averaging time
of 0.01 s, and correspondingly, their effects will average down by
more than a factor 100.  Similar considerations apply to the thermal
noise in the masses' suspension wires.  For detailed discussions of
fascinating and not-fully-understood physics issues that influence the
thermal noise, see, e.g., Refs.\ \cite{thermal1,thermal2,thermal3}.

\begin{figure}
\vskip 12.1pc
\special{hscale=53.5 vscale=53.5 hoffset=0 voffset=-52
psfile=seismicstack.eps}
\caption{The seismic isolation stack that was recently installed in
the LIGO Project's Mark II prototype interferometer at Caltech.
When the interferometer is in operation, a small tower is mounted on the
top steel plate and from the tower hangs
one of the interferometer's mirror-endowed
masses.  [Courtesy the LIGO Project.]
}
\label{fig:40mstack}
\end{figure}

At the LIGO sites, and most any other quiet location on Earth, the
ground is continually shaking with an rms displacement $\Delta l_{\rm
rms} \sim
10^{-8}\;{\rm cm}\;(100\;{\rm Hz}/f)^{3/2}$.  This is $10^7$ times
larger than the motions one seeks to monitor.  At frequencies above 10
Hz or so, one can protect the masses from these seismic vibrations by
simple (but carefully designed) passive isolation stacks.  Each
element in the stack is a mass and a spring (a harmonic oscillator) with
normal-mode frequency $f_0 \sim ($a few Hz).  When seismic noise
tries to drive this
harmonic oscillator far above its resonant frequency [in our case at
$f\agt ($a few tens of Hz)], the amplitude of its response is attenuated
relative to the driving
motion by a factor $(f_0/f)^2$ [in our case a factor $\agt 10^2$].
Thus, each oscillator in the stack will
provide a reduction $\agt 10^2$ in $\Delta l_{\rm rms}$, so a stack of four
or five
oscillators is enough to provide the required isolation.  Figure
\ref{fig:40mstack}
shows an isolation stack---made of four steel
plates and four sets of viton rubber
springs (not quite visible between the plates)---that is
now operating in the Mark II
prototype interferometer of Figure \ref{fig:mark2}.
This stack and the pendulum
wires that suspend the mirror-endowed test masses provide five layers
of isolation.  The installation of this new stack was responsible for the
sharp drop in
low-frequency noise in Figure \ref{fig:40mspectrum}
between June 1992 (Mark I) and March 1994 (Mark II).

The above rough estimates suggest (as Weiss realized as early as
1972 \cite{weiss})  that it is possible for interferometers
to achieve the required sensitivities, $h_{\rm rms} \sim 10^{-22}$ and
$\Delta L \sim 10^{-16}$ cm.  However, going from these rough
estimates to a real working interferometer, and doing so in the face of
a plethora of other noise sources, is a tremendous experimental
challenge---one that has occupied a number of excellent experimental
physicists since 1972.

\section{Resonant-Mass Antennas}
\label{resonantbars}

A resonant-mass antenna for gravitational radiation consists of a solid
body that (heuristically speaking) rings like a bell when a
gravitational wave hits it.  This body (the resonant mass) is usually a
cylinder, but future variants are likely to be spheres or sphere-like,
e.g.\ a truncated icosahedron gravitational-wave antenna or TIGA
\cite{tiga}.
The resonant mass is typically made from
an alloy of aluminum and weighs several tons, but some have been made of
niobium or single-crystal silicon or sapphire (but with masses well
below a ton).  To control thermal noise, the resonant mass is usually
cooled cryogenically to liquid-helium temperatures or below.

The resonant-mass antenna is instrumented with an electromagnetic transducer
and electronics, which monitor the complex amplitude of one or more of the
mass's normal modes.  When a gravitational wave passes through the mass,
its frequency components near each normal-mode frequency $f_o$
drive that mode, changing its complex amplitude;
and the time evolution of the changes is measured within some
bandwidth $\Delta f$ by the transducer and electronics.
Current resonant-mass antennas are narrow-band
devices ($\Delta f / f_o \ll 1$) but
in the era of LIGO/VIRGO, they might achieve bandwidths as large as
$\Delta f /f_o \sim 1/3$.

Resonant-mass antennas for gravitational radiation were pioneered by
Joseph Weber about 35 years ago \cite{weber}, and have been pushed to
ever higher sensitivity by Weber and a number of other research groups
since then.
For references and an overview of the present and future of such
antennas see, e.g., Ref.\ \cite{bassan}.
At present there is a network of such antennas \cite{pizella},
cooled to $3$K, and operating with an rms noise level for broad-band
gravity-wave bursts of $h_{\rm rms} \simeq 6 \times 10^{-19}$.  The
network includes an aluminum cylinder called EXPLORER
built by a group at the University of Rome, Italy
(Edoardo Amaldi, Guido Pizella, et.\ al.);
an aluminum cylinder at Louisiana State University, USA (Bill Hamilton,
Warren Johnson, et.\ al.); and a niobium cylinder at the University of Perth,
Australia (David Blair et.\ al.).  This network has been in operation,
searching for waves, for several years.

The next generation of resonant-mass antennas is now under
construction at the University of Rome (NAUTILUS) and at the
University of Legarno, Italy (AURIGA). These are several-ton
aluminum bars cooled to $0.05$K; their rms design sensitivities for
wave bursts are (several)$\times 10^{-20}$ \cite{bassan}.

A subsequent generation, which hopefully would operate in the
LIGO/VIRGO era, is being discussed and planned \cite{bassan}.
These are 1 to 100 ton
spheres or TIGA's cooled to $\sim 0.01$---$0.05$K, with
sensitivity goals of $\sim 10^{-21}$.  Such antennas might be built by
an American collaboration, a Brazilian collaboration, an Italian
collaboration called ``Omega'', and/or a Dutch collaboration called
``Grail''.  Their spherical or
near-spherical shapes make them omnidirectional and should give them
several-times higher sensitivities than can be achieved by cylinders
at the same frequency.

\begin{figure}
\vskip 12pc
\special{hscale=45 vscale=45 hoffset=-2 voffset=-9
psfile=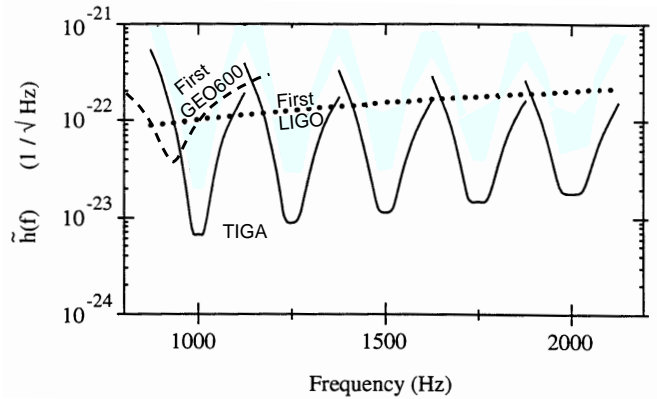}

\caption{
The rms noise curves $\tilde h(f)$ (measured in strain per root Hz) for
a ``xylaphone'' of TIGA gravitational-wave detectors (solid curves)
for signals of random polarization and direction \protect\cite{tiga}.  The
TIGA's are presumed instrumented and cooled sufficiently well that their
sensitivity is at the standard quantum limit.  Their central frequencies,
radii and masses (assuming aluminum material) are \{1.0 kHz, 1.30 m,
25.1 ton\}, \{1.25 kHz, 1.04 m, 12.8 ton\}, \{1.50 kHz, 0.87 m, 7.4
ton\}, \{1.75 kHz, 0.74 m, 4.7 ton\}, \{2.0 kHz, 0.65 m, 3.1
ton\}.  Shown for comparison are the noise curves for the first LIGO
interferometer with random wave polarization and direction
multiplied by $\protect\sqrt5$
(dotted curve; Fig.\ \protect\ref{fig:noisesources}),
and for the first GEO600 detector
operated in a
narrow-band mode (dashed curve; Ref.\ \protect\cite{geo}).
}
\label{fig:tiga}
\end{figure}

The attractiveness of such antennas in the LIGO/ VIRGO era lies in
their ability to operate with impressive sensitivity in the uppermost
reaches of the high-frequency band, $\sim 10^3$ to $10^4$ Hz, where
photon shot noise debilitates the performance of interferometric
detectors (cf.\ Fig.\ \ref{fig:noisesources}).
Figure \ref{fig:tiga} shows the projected rms
noise curves of a
family of TIGA detectors, each instrumented to operate at the
``standard quantum limit'' for such a detector (a nontrivial
experimental task).  Shown for comparison is the rms noise of the first LIGO
interferometer---which, of course, is not optimized for the kHz band.
The GEO600 interferometer, with its advanced design, can be operated in
a narrow-band, high-frequency mode (and probably will be so operated in
$\sim 1999$. Its rms design sensitivity in such a mode
is also shown in Figure \ref{fig:tiga}.  The TIGA sensitivities are
sufficiently
good in the kHz band, compared to early LIGO and GEO
interferometers, that, although they
probably cannot begin to operate until somewhat after the beginning of the
LIGO/VIRGO era, they might be fully competitive when they do
operate, and might play an important role in the kHz band.

\section{High-Frequency Gravitational-Wave Sources: Coalescing Compact
Binaries}
\label{coalescing_binaries}

The best understood of all gravitational-wave sources are coalescing,
compact binaries composed of neutron stars (NS) and black holes (BH).
These NS/NS, NS/BH, and BH/BH binaries may well become the ``bread and
butter'' of the LIGO/VIRGO diet.

The Hulse-Taylor \cite{hulse_taylor,taylor} binary pulsar,
PSR 1913+16, is an example of a NS/NS binary
whose waves could be measured by LIGO/VIRGO, if we were to wait long
enough.  At present PSR 1913+16 has an orbital frequency of about 1/(8 hours)
and emits its waves predominantly at twice this frequency, roughly
$10^{-4}$ Hz, which is in the low-frequency band---far too low to be
detected by LIGO/VIRGO.  However,
as a result of their loss of orbital energy to gravitational waves,
the PSR 1913+16 NS's are gradually spiraling inward.  If we wait roughly
$10^8$ years, this inspiral will bring the waves into the LIGO/VIRGO
high-frequency band. As the NS's continue their inspiral,
the waves will then sweep upward in frequency, over a time of about 15
minutes, from 10 Hz to $\sim 10^3$ Hz, at which point the NS's will
collide and coalesce.  It is this last 15 minutes
of inspiral, with $\sim 16,000$
cycles of waveform oscillation, and the final coalescence,
that LIGO/VIRGO seeks to monitor.

\begin{figure}
\center
\vskip16.8pc
\special{hscale=43 vscale=43 hoffset=3 voffset=-5
psfile=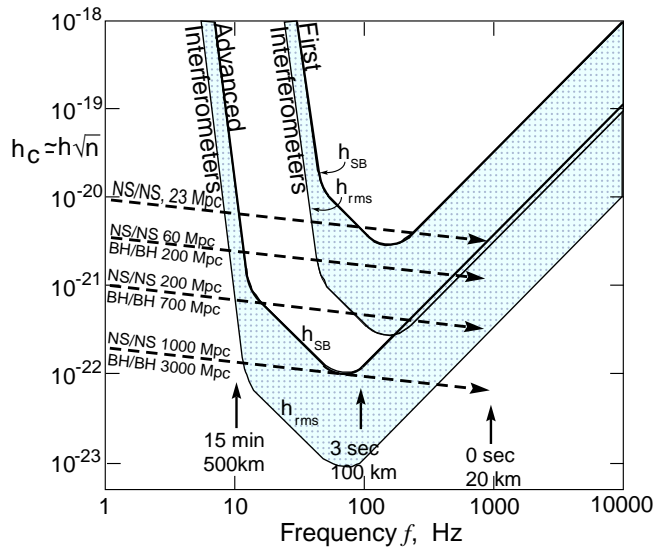}
\caption{LIGO's projected broad-band noise $h_{\rm rms}$ and sensitivity
to bursts $h_{\rm SB}$ (Fig.\ \protect\ref{fig:noisesources} and Ref.\
\protect\cite{ligoscience}) compared with the
strengths of the
waves from the last few minutes of inspiral of compact binaries.  The
signal to noise ratios are
$\protect\sqrt 2$ higher than in Ref.\ \protect\cite{ligoscience}
because of a factor 2 error in Eq.~(29) of Ref.\ \protect\cite{300yrs}.
}
\label{fig:CBStrengthSensitivity}
\end{figure}

\subsection{Wave Strengths Compared to LIGO Sensitivities}
\label{cbsensitivities}

Figure \ref{fig:CBStrengthSensitivity}
compares the projected sensitivities of interferometers in
LIGO \cite{ligoscience}
with the wave strengths from the last few minutes of inspiral
of BH/BH, NS/BH, and NS/NS binaries at various distances from Earth.  The
two solid
curves at the bottoms of the stippled regions (labeled $h_{\rm rms}$)
are the rms noise levels
for broad-band waves that have optimal direction and polarization.
The tops of the stippled regions (labeled $h_{\rm SB}$ for ``sensitivity to
bursts'') are the sensitivities for highly confident
detection of randomly polarized, broad-band waves from random directions
(i.e., the sensitivities for high confidence that any such observed
signal is not a false alarm due to Gaussian noise).  The upper stippled
region and its bounding curves
are the expected performances of the first interferometers in LIGO; the
lower stippled region and curves are performances of more advanced
LIGO interferometers; cf.\ Figure \ref{fig:noisesources}.

As the NS's and/or BH's spiral inward, their waves
sweep upward in frequency (left to right in the diagram).  The dashed
lines show their ``characteristic'' signal strength $h_c$
(approximately the amplitude $h$ of the waves' oscillations
multiplied by the square root of the number of cycles spent near a given
frequency, $\sqrt n$); the signal-to-noise ratio is this $h_c$
divided by the detector's $\sqrt5 h_{\rm rms}$, $S/N = h_c/(\sqrt5
h_{\rm rms})$, where the $\sqrt5$
converts $h_{\rm rms}$ from ``optimal direction and polarization'' to
``random direction and polarization'') \cite{ligoscience,300yrs}.  The arrows
along the bottom inspiral track indicate the time until
final coalescence for an NS/NS binary and the separation between the
NS centers
of mass.  Each NS is assumed to have a mass of 1.4 suns and a radius
$\sim 10$ km; each BH, 10 suns and $\sim 20$ km.

Notice that the signal strengths in Figure \ref{fig:CBStrengthSensitivity}
are in good accord with our rough estimates based on Eq.\ (\ref{hom});
at the endpoint (right end) of each inspiral, the number of cycles $n$ spent
near that frequency is of order unity, so the quantity plotted, $h_c
\simeq h\sqrt
n$, is about equal to $h$---and at distance 200 Mpc is roughly $10^{-21}$,
as we estimated in Section \ref{armlength}.

\subsection{Coalescence Rates}
\label{coalescencerates}

Such final coalescences are few and far between in our own galaxy:
about one every 100,000 years, according to 1991
estimates by Phinney \cite{phinney} and
by Narayan, Piran, and Shemi \cite{narayan},
based on the statistics of binary pulsar searches
in our galaxy which found three that will coalesce in less than
$10^{10}$ years.  Extrapolating out through the universe
on the basis of the density of
production of blue light (the color produced predominantly by
massive stars), Phinney \cite{phinney} and Narayan et.\ al.\ \cite{narayan}
infer that to see several
NS/NS coalescences per year, LIGO/VIRGO will have to look out to a
distance of about 200 Mpc (give or take a factor $\sim 2$); cf.\ the
``NS/NS inspiral, 200 Mpc'' line in Figure \ref{fig:CBStrengthSensitivity}.
Since these estimates were made, the binary pulsar searches have been
extended through a significantly larger volume of the galaxy than
before, and no new ones with coalescence times $\alt 10^{10}$ years have
been found; as a result, the binary-pulsar-search-based
best estimate of the coalescence rate should be
revised downward \cite{bailes}, perhaps to as little as one every
million years in our
galaxy, corresponding to a distance $400$ Mpc for several per year
\cite{bailes}.

A rate of one every million years
in our galaxy is $\sim 1000$ times smaller than the birth rate of
the NS/NS binaries' progenitors:
massive, compact, main-sequence binaries \cite{phinney,narayan}.
Therefore, either 99.9 per cent of progenitors
fail to make it to the NS/NS state (e.g., because of binary disruption
during a supernova or forming T\.ZO's),
or else they do make it, but they wind up as a
class of NS/NS binaries that has not yet been discovered in any of the
pulsar searches.  Several experts on binary evolution have argued for
the latter \cite{tutukov_yungelson,yamaoka,lipunov}: most
NS/NS binaries, they suggest, may form with such short orbital periods
that their lifetimes to coalescence are significantly shorter than
normal pulsar lifetimes ($\sim 10^7$ years); and with such short
lifetimes, they have been missed in pulsar searches.  By modeling the
evolution of the galaxy's binary star population, the binary experts arrive at
best estimates as high as $3\times 10^{-4}$ coalescences per year in our
galaxy, corresponding to several per year out to 60 Mpc distance
\cite{tutukov_yungelson}.  Phinney \cite{phinney} describes other
plausible populations of NS/NS binaries that could increase the event
rate, and he argues for ``ultraconservative'' lower and upper limits of
23 Mpc and 1000Mpc
for how far one must look to see several coalescence per year.

By comparing these rate estimates with the signal strengths in Figure
\ref{fig:CBStrengthSensitivity}, we see that (i) the first interferometers
in LIGO/VIRGO have a possibility but not high probability of seeing
NS/NS coalescences; (ii) advanced interferometers are almost certain of
seeing them (the requirement that
this be so was one factor that forced the LIGO/VIRGO arm lengths to be
so long, several kilometers); and (iii) they are most likely to be
discovered roughly half-way between the first and advanced
interferometers---which means by an improved variant of the first
interferometers several years after LIGO operations begin.

We have no good observational handle on the coalescence rate of NS/BH or
BH/BH binaries.  However, theory suggests that their progenitors might
not disrupt during the stellar collapses that produce the NS's and BH's,
so their coalescence rate could be about the same as the birth
rate for their progenitors: $\sim 1/100,000$ years in our galaxy.  This
suggests that within 200 Mpc distance there might be several NS/BH or
BH/BH coalescences per year.
\cite{phinney,narayan,tutukov_yungelson,lipunov}.
This estimate should be regarded as a
plausible upper limit on the event rate and lower limit on the distance to
look \cite{phinney,narayan}.

If this estimate is correct, then NS/BH and BH/BH binaries will be seen
before NS/NS, and might be seen by the first LIGO/VIRGO interferometers
or soon thereafter; cf.\ Figure  \ref{fig:CBStrengthSensitivity}.
However, this estimate is far less certain than the
(rather uncertain) NS/NS estimates!

Once coalescence waves have been discovered, each further improvement of
sensitivity by a factor 2 will increase the event rate by $2^3 \simeq
10$.  Assuming a rate of several NS/NS per year at 200 Mpc, the advanced
interferometers of Figure \ref{fig:CBStrengthSensitivity} should see
$\sim 100$ per year.

\subsection{Inspiral Waveforms and the Information They Can Bring}
\label{cbwaveforms}

Neutron stars and black holes have such intense self gravity that it is
exceedingly difficult to deform them.  Correspondingly, as they spiral
inward in a compact binary, they do not gravitationally deform each other
significantly until several orbits before their final
coalescence \cite {kochanek,bildsten_cutler}.  This means
that the inspiral waveforms are determined to high accuracy by
only a few, clean parameters:
the masses and spin angular momenta of the bodies, and the initial
orbital elements (i.e.\ the elements when the waves enter the LIGO/VIRGO band).

\begin{figure}
\vskip 10.1pc
\special{hscale=45 vscale=45 hoffset=-8 voffset=-13
psfile=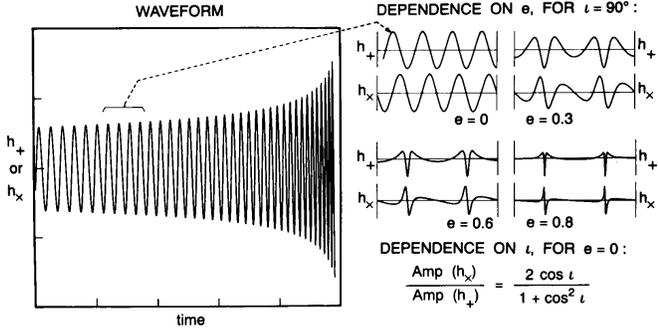}
\caption{Waveforms from the inspiral of a compact binary, computed using
Newtonian gravity for the orbital evolution and the quadrupole-moment
approximation for the wave generation.  (From Ref.\
\protect\cite{ligoscience}.)}
\label{fig:NewtonInspiral}
\end{figure}

Though tidal deformations are negligible during inspiral, relativistic
effects can be very important.
If, for the moment, we ignore the relativistic effects---i.e., if we
approximate gravity as Newtonian and the wave generation as due to the
binary's oscillating quadrupole moment \cite{300yrs},
then the shapes of the inspiral
waveforms $h_+(t)$ and $h_\times(t)$
are as shown in Figure \ref{fig:NewtonInspiral}.

The left-hand graph in Figure \ref{fig:NewtonInspiral}
shows the waveform increasing in
amplitude and sweeping upward in frequency
(i.e., undergoing a ``chirp'')
as the binary's bodies spiral closer and closer together.  The ratio of
the amplitudes
of the two polarizations is determined by the inclination $\iota$ of the
orbit to our line of sight (lower right in Fig.\ \ref{fig:NewtonInspiral}).
The
shapes of the individual waves, i.e.\ the waves' harmonic content, are
determined by the orbital eccentricity (upper right).  (Binaries
produced by normal stellar evolution should be highly circular due to
past radiation reaction forces, but compact
binaries that form by capture events, in dense star clusters that might
reside in galactic nuclei \cite{quinlan_shapiro}, could be quite
eccentric.)  If, for simplicity, the
orbit is circular, then the rate at which
the frequency sweeps or ``chirps'', $df/dt$
[or equivalently the number of cycles
spent near a given frequency, $n=f^2(df/dt)^{-1}$] is determined solely, in the
Newtonian/quadrupole approximation, by the binary's so-called {\it
chirp mass}, $M_c \equiv (M_1M_2)^{3/5}/(M_1+M_2)^{1/5}$ (where $M_1$
and $M_2$ are the two bodies' masses).
The amplitudes of the two waveforms are determined by the chirp mass,
the distance to the source, and the orbital inclination.  Thus
(in the Newtonian/quadrupole
approximation), by measuring the two amplitudes, the frequency sweep, and
the harmonic content of the inspiral waves, one can determine as direct,
resulting observables, the source's distance, chirp mass, inclination,
and eccentricity \cite{schutz_nature86,schutz_grg89}.

\begin{figure}
\vskip12.6pc
\special{hscale=100 vscale=100 hoffset=-10 voffset=-10
psfile=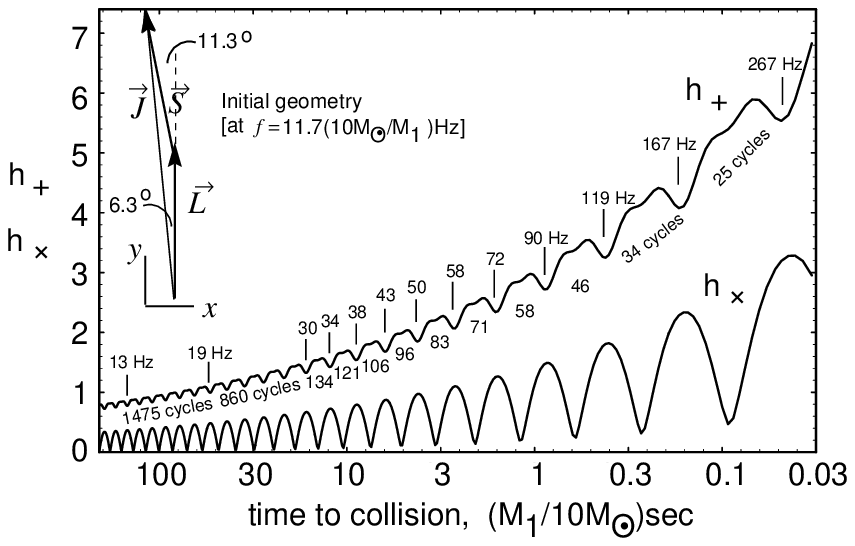}
\caption{Modulational envelope for the waveform from
a $1 M_\odot$ nonspinning NS spiraling into a $10 M_\odot$, rapidly
spinning Kerr black hole (spin parameter $a = 1$).  The orbital angular
momentum $\bf L$ is inclined by $\alpha = 11.3$ degrees to the hole's
spin angular momentum $\bf S$, and the two precess around
${\bf J} ={\bf L} + {\bf S}$, whose direction remains
fixed in space as $L = |{\bf L}|$ shrinks and $S = |{\bf S}| = M_{\rm
BH}a$ remains constant.
The precession
modulates the waves by an amount that depends on
(i) the direction to Earth (here along the
initial $\bf L \times \bf S$, i.e. out of the paper)
and (ii) the
orientation of the detector's arms (here parallel to the figure's
initial $\bf L$ and to ${\bf L}\times$(direction to Earth)
for $h_+$, and rotated 45 degrees for $h_\times$).
The figure shows the waveforms' modulational envelopes (in
arbitrary units, the same for $h_+$ and $h_\times$),
parametrized by the wave frequency $f$ and the number of cycles
of {\it oscillation} between the indicated $f$'s.  The total
number of {\it precessions} from
$f$ to coalescence is $N_{\rm prec} \simeq (5/64\pi)(Ma/\mu)(\pi
Mf)^{-2/3} \simeq 20(f/10\hbox{Hz})^{-2/3}$.
(From \protect\cite{last3minutes,precess}.)
}
\label{fig:precess}
\end{figure}

As in binary pulsar observations \cite{taylor},
so also here, relativistic effects add further information:
they influence the rate of frequency sweep and produce waveform
modulations in ways that
depend on the binary's dimensionless ratio $\eta = \mu/M$ of reduced mass
$\mu = M_1 M_2/(M_1 + M_2)$ to total mass $M = M_1 + M_2$ \cite{lincoln_will}
and on the spins of the binary's two bodies \cite{kidder_will_wiseman}.
These relativistic effects are reviewed and discussed at length in
Refs.\ \cite{last3minutes,will_nishinomiya}.  Two deserve special
mention: (i) As the waves emerge from the binary, some of them get
backscattered one or more times off the binary's spacetime curvature,
producing wave {\it tails}.  These tails act back on the binary,
modifying its inspiral rate in a measurable way.  (ii)  If the orbital
plane is inclined to one or both of the binary's spins, then
the spins drag inertial frames in the binary's
vicinity (the ``Lense-Thirring effect''), this frame dragging causes
the orbit to precess, and the precession modulates the
waveforms \cite{last3minutes,precess,kidder}.  Figure
\ref{fig:precess} shows the resulting modulation for a $1 M_\odot$
NS spiraling into a rapidly spinning, $10 M_\odot$ BH.

Remarkably, the relativistic corrections to the
frequency sweep will be measurable with very high
accuracy, even though they are typically $\alt 10$ per cent of the
Newtonian contribution, and even though the
typical signal to noise ratio will be only $\sim 9$ even after optimal
signal processing.  The reason is as
follows \cite{cutler_flanagan,finn_chernoff,last3minutes}:

The frequency sweep will be monitored by the method of ``matched
filters''; in other words, the incoming, noisy signal will be cross
correlated with theoretical templates.  If the signal and the templates
gradually
get out of phase with each other by more than $\sim 1/10$ cycle as the
waves sweep
through the LIGO/VIRGO band, their cross correlation will be significantly
reduced.
Since the total number of cycles spent in the LIGO/VIRGO band will
be $\sim 16,000$ for a NS/NS binary, $\sim 3500$ for NS/BH, and $\sim
600$ for BH/BH, this means that LIGO/VIRGO should be able to measure the
frequency sweep to a fractional precision $\alt 10^{-4}$,
compared to which the relativistic effects
are very large.  (This is essentially the same method as
Joseph Taylor and colleagues use for high-accuracy radio-wave measurements of
relativistic effects in binary pulsars \cite{taylor}.)

Preliminary analyses, using the theory of optimal signal processing,
predict the following typical accuracies for LIGO/VIRGO measurements
based solely on the frequency sweep (i.e., ignoring modulational
information)
\cite{poisson_will,cutler_flanagan,finn_chernoff},
\cite{jaronowski_krolak,last3minutes}: (i) The
chirp mass $M_c$
will typically be measured, from the Newtonian part of the frequency
sweep, to $\sim 0.04\%$ for a NS/NS binary and
$\sim 0.3\%$ for a system containing at least one BH.
(ii) {\it If} we
are confident (e.g., on a statistical basis from measurements of many
previous binaries) that the spins are a few percent or less
of the maximum physically allowed, then the reduced mass $\mu$
will be measured to
$\sim 1\%$ for NS/NS and NS/BH binaries, and
$\sim 3\%$ for BH/BH binaries.  (Here and below NS means a
$\sim 1.4 M_\odot$
neutron star and BH means a $\sim 10 M_\odot$
black hole.) (iii) Because the
frequency dependences
of the (relativistic) $\mu$ effects
and spin effects are not
sufficiently different
to give a clean separation between $\mu$ and the spins,
if we have no prior knowledge of the spins, then
the spin$/\mu$ correlation will
worsen the typical accuracy of $\mu$ by a large factor,
to $\sim 30\%$ for NS/NS, $\sim 50\%$ for NS/BH, and
a factor $\sim 2$ for BH/BH \cite{poisson_will,cutler_flanagan}.
These worsened accuracies might be improved somewhat
by waveform modulations caused by the
spin-induced precession of the orbit \cite{precess,kidder},
and even without modulational information, a certain
combination of $\mu$ and the spins
will be determined to a few per cent.  Much
additional theoretical work is needed
to firm up the measurement accuracies.

To take full advantage of all the information in the inspiral waveforms
will require theoretical templates that are accurate, for given masses
and spins, to a fraction of a cycle during the entire sweep through the
LIGO/VIRGO band.  Such templates are being computed by an international
consortium of relativity theorists (Blanchet and Damour in France, Iyer
in India, Will and Wiseman in the U.S., and others)
\cite{will_nishinomiya,2pnresults}, using post-Newtonian expansions of
the Einstein field equations.  This enterprise is rather like computing
the Lamb shift to high order in powers of the fine structure
constant, for comparison with experiment.  The terms of leading order in
the mass ratio $\eta = \mu / M$ are being checked by a
Japanese-American consortium
(Nakamura, Sasaki, Tagoshi, Tanaka, Poisson) using the Teukolsky
formalism for weak perturbations of black holes
\cite{poisson_check,shibataetal}.  These small-$\eta$ calculations have
been carried to very high post-Newtonian order for circular orbits and no spins
\cite{nakamura_tagoshi,sasaki_tagoshi}, and from those results Cutler
and Flanagan \cite{cutler_flanagan1} have estimated the order to which the
full, finite-$\eta$ computations must be carried in order that
systematic errors in the theoretical templates will not significantly
impact the information extracted from the LIGO/VIRGO observational data.
The answer appears daunting: radiation-reaction effects must be computed
to three full post-Newtonian orders [six orders in $v/c =$(orbital
velocity)/(speed of light)] beyond the leading-order radiation reaction,
which itself is 5 orders in $v/c$ beyond the Newtonian theory of
gravity.

It is only about ten years since controversies over the leading-order
radiation reaction \cite{quadrupole_controversy} were resolved by a
combination of theoretical
techniques and binary pulsar observations.  Nobody dreamed then that
LIGO/VIRGO observations will require pushing post-Newtonian computations
onward from $O[(v/c)^5]$ to $O[(v/c)^{11}]$.  This requirement epitomizes
a major change in the field of relativity research: At last, 80 years
after Einstein formulated general relativity, experiment has become a
major driver for theoretical analyses.

Remarkably, the goal of $O[(v/c)^{11}]$ is achievable.  The most difficult
part of the computation, the radiation reaction, has been evaluated to
$O[(v/c)^9]$ beyond Newton by the French/Indian/American consortium
\cite{2pnresults}
and as of this writing, rumors have it that $O[(v/c)^{10}]$ is coming
under control.

These high-accuracy waveforms are needed only for extracting information
from the inspiral waves, after the waves have been discovered; they are
not needed for the discovery itself.  The discovery is best achieved
using a different family of theoretical waveform templates, one that
covers the space of potential waveforms
in a manner that minimizes computation time instead
of a manner that ties quantitatively into general relativity
theory \cite{last3minutes}.  Such templates are in the early stage of
development \cite{sathyaprakash,krolak_kokkotas_schafer,apostolatos}.

LIGO/VIRGO observations of compact binary inspiral have the potential to
bring us far more information than just binary masses and spins:
\begin{itemize}
\item
They can be used for high-precision tests of general relativity.  In
scalar-tensor theories (some of which are attractive alternatives
to general relativity \cite{damour_nordvedt}), radiation reaction
due to emission
of scalar waves places a unique signature on the gravitational
waves that LIGO/VIRGO
would detect---a signature that can be searched for with high precision
\cite{will_scalartensor}.
\item
They can be used to measure the Universe's Hubble constant, deceleration
parameter, and cosmological constant
\cite{schutz_nature86,schutz_grg89,markovic,chernoff_finn}.  The keys to
such measurements are that: (i) Advanced interferometers in
LIGO/VIRGO will be able to see NS/NS
out to cosmological redshifts $z \sim 0.3$, and NS/BH out to $z
\sim 2$. (ii) The direct observables that can be extracted
from the
observed waves include the source's luminosity distance $r_{\rm L}$ (measured
to accuracy $\sim 10$ per cent in a large fraction of cases), and its
direction on the sky (to accuracy $\sim 1$ square degree)---accuracies
good enough that only one or a few electromagnetically-observed
clusters of galaxies should fall within the 3-dimensional
gravitational error boxes, thereby giving promise to joint
gravitational/electromagnetic statistical studies.  (iii) Another direct
gravitational observable is $(1+z)M$
where $z$ is redshift and $M$ is any mass in the system (measured to the
accuracies quoted above). Since the masses of NS's in binaries seem to
cluster around $1.4 M_\odot$, measurements of $(1+z)M$ can provide a
handle on the redshift, even in the absence of electromagnetic aid.
\item
For a NS or small BH spiraling into a massive $\sim 50$ to $500 M_\odot$
BH, the inspiral waves will carry a ``map'' of the spacetime geometry
around the big hole---a map that can be used, e.g., to test the theorem
that ``a black hole has no hair'' \cite{ryan_finn_thorne};
cf.\ Section \ref{lfbhinspiral} below.
\end{itemize}

\subsection{Coalescence Waveforms and their Information}
\label{coalescence_waves}

The waves from the binary's final coalescence can bring us new
types of information.

\bigskip
\centerline{\it BH/BH Coalescence}
\medskip

In the case of a BH/BH binary, the coalescence
will excite large-amplitude, highly nonlinear vibrations of spacetime
curvature near the coalescing black-hole horizons---a phenomenon of
which we have very little theoretical understanding today.  Especially
fascinating will be the case of two spinning black holes whose spins are
not aligned with each other or with the orbital angular momentum.  Each
of the three angular momentum vectors (two spins, one orbital) will drag
space in its vicinity into a tornado-like swirling motion---the general
relativistic ``dragging of inertial frames,'' so the binary is rather
like two tornados with orientations skewed to each other, embedded inside a
third, larger tornado with a third orientation.  The dynamical evolution of
such a complex configuration of coalescing spacetime warpage
(as revealed by its
emitted waves) might bring us
surprising new insights into relativistic gravity \cite{ligoscience}.
Moreover, if the sum
of the BH masses is fairly large, $\sim 40$ to $200 M_\odot$, then the waves
should come off in a frequency range $f\sim 40$ to $200$ Hz where the
LIGO/VIRGO broad-band interferometers have their best sensitivity and can best
extract
the information the waves carry.

To get full value out of such wave observations will require
\cite{hughes_flanagan} having
theoretical computations with which to compare them.
There is no hope to perform such computations
analytically; they can only be done as supercomputer simulations.
The development of such simulations
is being pursued by several research groups, including an
eight-university American consortium of numerical relativists and
computer scientists called the
Two-Black-Hole Grand Challenge Alliance \cite{GC} (Co-PIs:
Richard Matzner and Jim Browne, U.\ Texas Austin;
Larry Smarr, Ed Seidel, Paul Saylor, Faisal Saied, U.\ Illinois Urbana;
Geoffrey Fox, Syracuse U.;
Stu Shapiro and Saul Teukolsky, Cornell U.;
Jim York and Charles Evans, U.\ North Carolina;
Sam Finn, Northwestern U.;
Pablo Laguna, Pennsylvania State U.;
and Jeff Winicour, U. Pittsburgh).  I have a bet with
Matzner, the lead PI of this alliance, that LIGO/VIRGO will discover
waves from such coalescences with misaligned spins before the Alliance
is able to compute them.

\bigskip
\centerline{\it NS/NS Coalescence}
\medskip

The final coalescence of NS/NS binaries should produce waves that are
sensitive to the equation of state of nuclear matter, so
such coalescences have the potential to teach us about the
nuclear equation of state \cite{ligoscience,last3minutes}.  In essence,
LIGO/VIRGO will be
studying nuclear physics via the collisions of atomic nuclei that have
nucleon numbers $A \sim 10^{57}$---somewhat larger than physicists are normally
accustomed to.
The accelerator used to drive these nuclei up to the speed of light is
the binary's self gravity, and the radiation by which the details of the
collisions are probed is gravitational.

Unfortunately, the final NS/NS coalescence will emit its gravitational
waves in the kHz frequency band ($800 {\rm Hz} \alt f \alt 2500 {\rm
Hz}$) where photon shot noise will prevent them from being studied by
the standard, ``workhorse,'' broad-band
interferometers of Figure \ref{fig:noisesources}.
However, a specially configured (``dual-recycled'')
interferometer invented by Brian Meers \cite{meers},
which could have enhanced sensitivity in the kHz
region at the price of reduced sensitivity elsewhere, may be able to
measure the waves and extract their equation of state information,
as might massive, spherical, resonant-mass detectors
\cite{last3minutes,kennefick_laurence_thorne}. Such measurements will
be very difficult and are likely only when the LIGO/VIRGO
network has reached a mature stage.

A number of research groups
\cite{nakamura_etal_nsns2,rasio_shapiro,nakamura_nishinomiya},
\cite{centrella,davies_melvyn}
are engaged in numerical astrophysics
simulations of NS/NS coalescence, with the goal not only to predict the
emitted gravitational waveforms and their dependence on equation of
state, but also (more immediately) to learn whether such
coalescences
might power the $\gamma$-ray bursts that have been a major astronomical
puzzle since their discovery in the early 1970s.

NS/NS coalescence is
currently a popular explanation for the $\gamma$-ray bursts because
(i) the bursts are isotropically distributed on the sky, (ii) they have
a distribution of number versus intensity that suggests they might lie
at near-cosmological distances, and (iii) their event rate is
roughly the same as that predicted for NS/NS coalescence ($\sim1000$
per year out to cosmological distances, if they are cosmological).
If LIGO/VIRGO were now in operation and observing
NS/NS inspiral, it could
report definitively whether or not the $\gamma$-bursts are produced by
NS/NS binaries; and if the answer were yes, then the combination of
$\gamma$-burst data and gravitational-wave data could bring valuable
information that neither could bring by itself.  For example, it would
reveal when, to within a few msec, the $\gamma$-burst is emitted
relative to the moment the NS's first begin to touch; and by
comparing the $\gamma$ and gravitational times of arrival,
we might test whether gravitational waves propagate with
the speed of light to a fractional precision of
$\sim 0.01{\rm sec}/3\times10^9\, {\rm lyr} = 10^{-19}$.

\bigskip
\centerline{\it NS/BH Coalescence}
\medskip

A NS spiraling into a BH of mass $M \agt 10
M_\odot$ should be swallowed more or less whole.  However, if the BH is
less massive than roughly $10 M_\odot$, and especially if it is rapidly
rotating, then the NS will tidally disrupt before being swallowed.
Little is known about the disruption and accompanying waveforms.  To
model them with any reliability will likely require full numerical
relativity, since the circumferences of the BH and NS will be comparable
and their physical separation at the moment of disruption
will be of order their separation. As with NS/NS, the coalescence
waves should
carry equation of state information and will come out in the kHz band,
where their detection will require advanced, specialty detectors.

\bigskip
\centerline{\it Christodoulou Memory}
\medskip

As the coalescence waves depart from
their source, their energy creates (via the nonlinearity of Einstein's
field equations) a secondary wave called the ``Christodoulou memory''
\cite{christodoulou,thorne_memory,wiseman_will_memory}.  Whereas the primary
waves may have frequencies in the kHz band, the memory builds up on the
timescale of the primary energy emission profile, which is likely to be
of order 0.01 sec, corresponding to a memory frequency in the optimal
band for the LIGO/VIRGO workhorse interferometers, $\sim 100$Hz.
Unfortunately, the memory is so weak that only very advanced
interferometers have much chance of detecting and studying it---and
then, perhaps only for BH/BH coalescences and not for NS/NS or NS/BH
\cite{kennefick_memory}.

\section{Other High-Frequency Sources}
\label{otherhfsources}

\subsection{Stellar Core Collapse and Supernovae}
\label{supernovae}

When the core of a massive star has exhausted its supply of nuclear fuel,
it collapses to form a neutron star or black hole. In some cases, the
collapse triggers and powers a subsequent explosion
of the star's mantle---a supernova explosion.  Despite extensive
theoretical efforts for more than 30 years, and despite wonderful
observational data from Supernova 1987A, theorists are still far from
a definitive understanding of the details of the collapse and explosion.  The
details are highly complex and may differ greatly from one
core collapse to another \cite{petschek}.

Several features of the collapse and the core's subsequent
evolution can produce significant gravitational radiation in the
high-frequency band. We shall
consider these features in turn, the most weakly radiating first.

\bigskip
\centerline{\it Boiling of the Newborn Neutron Star}
\medskip

\begin{figure}
\vskip16.8pc
\special{hscale=43 vscale=43 hoffset=1 voffset=-5
psfile=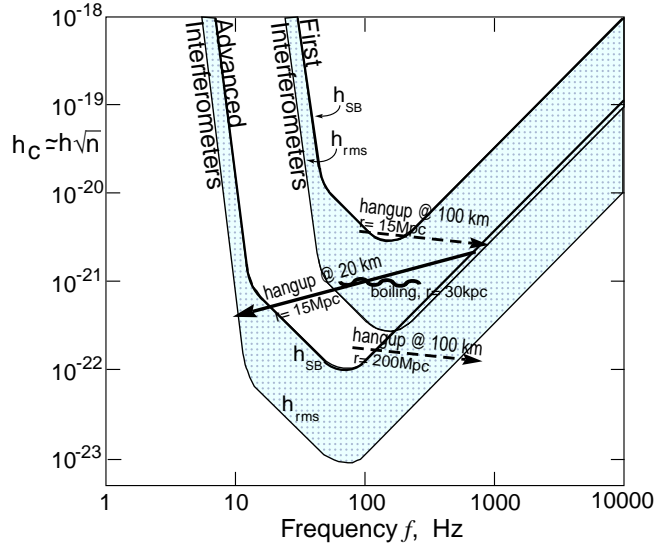}
\caption{Characteristic amplitudes of the gravitational waves from
various processes accompanying stellar core collapse and supernovae,
compared with projected sensitivities of LIGO's interferometers.
}
\label{fig:supernovae}
\end{figure}

Even if the collapse is spherical, so it cannot radiate any
gravitational waves at all, it should
produce a convectively unstable neutron
star that ``boils'' vigorously (and nonspherically) for the first
$\sim 0.1$ second of its life \cite{bethe}.  The boiling dredges
up high-temperature
nuclear matter ($T\sim 10^{12}$K) from the neutron star's central regions,
bringing it to the surface (to the ``neutrino-sphere''), where it
cools by
neutrino emission before being swept back downward and reheated.  Burrows
estimates \cite{burrows,burrows1} that the  boiling
should generate $n \sim 10$ cycles of gravitational waves with
frequency $f\sim 100$Hz and amplitude
$h \sim 3 \times 10^{-22} (30{\rm kpc}/r)$ (where $r$ is the distance to
the source), corresponding to a characteristic amplitude $h_c \simeq
h\sqrt n \sim 10^{-21} (30{\rm kpc}/r)$; cf.\ Figure
\ref{fig:supernovae}.  LIGO/VIRGO will be able to detect such waves only
in the local group of galaxies, where the supernova rate is probably no
larger than $\sim 1$ each 10 years.  However, neutrino detectors have a
similar range, and there could be a high scientific payoff from
correlated observations of the gravitational waves emitted by the
boiling's mass motions and neutrinos emitted from the boiling
neutrino-sphere.

\bigskip
\centerline{\it Axisymmetric Collapse, Bounce, and Oscillations}
\medskip

Rotation will centrifugally flatten the collapsing
core, enabling it to radiate as it implodes.  If the core's angular
momentum is small enough that centrifugal forces do
not halt or strongly slow the collapse before it reaches
nuclear densities, then the core's collapse, bounce, and subsequent
oscillations are likely to be axially symmetric.  Numerical
simulations \cite{finn_collapse,monchmeyer} show that in this case the waves
from collapse, bounce, and oscillation
will be quite weak: the total energy radiated as gravitational waves
is not likely to exceed $\sim 10^{-7}$ solar masses (about 1 part in a
million of the collapse energy) and might often be much less than
this; and correspondingly, the waves'
characteristic amplitude will be $h_c \alt 3\times 10^{-21}(30{\rm
kpc}/r)$.  These collapse-and-bounce waves will come off at frequencies
$\sim 200$ Hz to $\sim 1000$ Hz, and will precede the boiling waves by a
fraction of a second.  Like the boiling waves, they probably cannot be
seen by LIGO/VIRGO beyond the local group of galaxies and thus will be a
very rare occurrence.

\bigskip
\centerline{\it Rotation-Induced Bars and Break-Up}
\medskip

If the core's rotation is large enough to strongly flatten the
core before or as it reaches nuclear density,
then a dynamical and/or
secular instability is likely to break the core's axisymmetry.
The core will be transformed into
a bar-like configuration that spins end-over-end like
an American football, and that might even break up into two or more massive
pieces.  In this case, the radiation from the spinning bar or orbiting
pieces {\it could} be almost as strong as that from a coalescing neutron-star
binary, and thus could be seen by the LIGO/VIRGO first interferometers
out to the distance of the Virgo cluster (where the supernova rate is
several per
year) and by advanced interferometers out to several hundred Mpc
(supernova rate $\sim 10^4$ per year); cf.\ Figure \ref{fig:supernovae}.
It is far
from clear what fraction of collapsing cores will have enough angular
momentum to break their axisymmetry, and what fraction of those will
actually radiate at this high rate; but even if only $\sim 1/1000$ or
$1/10^4$ do so, this could ultimately be a very interesting source for
LIGO/VIRGO.

Several specific scenarios for such non-axisymmetry have been identified:

{\bf Centrifugal hangup at $\bf \sim 100$km radius:} If the
pre-collapse core is rapidly spinning (e.g., if it is a white dwarf that
has been spun up by accretion from a companion), then the collapse may
produce a highly flattened, centrifugally supported disk with most of
its mass at radii $R\sim 100$km, which then (via instability)
may transform itself into a bar or may bifurcate.  The bar or
bifurcated lumps will radiate gravitational waves at twice their rotation
frequency, $f\sim 100$Hz --- the optimal frequency for LIGO/VIRGO
interferometers.  To shrink on down to $\sim 10$km size, this
configuration must shed most of its angular momentum.  {\it If} a
substantial fraction of the angular momentum goes into
gravitational waves, then independently of the strength of the bar,
the waves will be nearly as strong as those from a coalescing binary.
The reason is this:
The waves' amplitude $h$ is proportional to the bar's ellipticity $e$,
the number of cycles $n$ of wave emission is proportional to $1/e^2$, and the
characteristic amplitude $h_c = h\sqrt n$ is thus independent of the
ellipticity and is about the same whether the configuration is a bar or
is two lumps \cite{schutz_grg89}.  The resulting waves will thus have $h_c$
roughly half as large, at $f\sim 100$Hz, as the $h_c$ from a NS/NS binary
(half as large because each lump might be half as massive as a NS), and
the waves will chirp upward in frequency in a manner similar to those from a
binary.

It is rather likely, however, that most of excess angular momentum does {\it
not} go into gravitational waves, but instead goes largely into hydrodynamic
waves as the bar or lumps, acting like a propeller, stir up the
surrounding stellar mantle.  In this case, the radiation will be
correspondingly weaker.

{\bf Centrifugal hangup at $\bf \sim 20$km radius:}  Lai and Shapiro
\cite{lai} have explored the case of centrifugal
hangup at radii not much larger than the final neutron star, say $R\sim
20$km.  Using compressible ellipsoidal models, they have deduced that,
after a brief period of dynamical bar-mode instability with wave
emission at $f\sim 1000$Hz (explored by
Houser, Centrella, and Smith \cite{houser}), the star switches to a secular
instability in which the bar's angular velocity gradually slows while
the material of which it is made retains its high rotation speed and
circulates through the slowing bar.  The slowing bar emits waves that sweep
{\it downward} in frequency through the LIGO/VIRGO optimal band $f\sim 100$Hz,
toward $\sim 10$Hz. The characteristic amplitude (Fig.\
\ref{fig:supernovae}) is only modestly smaller than for the upward-sweeping
waves from hangup at $R\sim 100$km, and thus such waves should be
detectable near the Virgo Cluster by the first LIGO/VIRGO interferometers,
and at distances of a few 100Mpc by advanced interferometers.

{\bf Successive fragmentations of an accreting, newborn neutron star:}
Bonnell and Pringle \cite{pringle} have focused on the evolution of the
rapidly spinning, newborn neutron star as it quickly accretes more and
more mass from the pre-supernova star's inner mantle.  If the accreting
material carries high angular momentum, it may trigger a renewed bar
formation, lump formation, wave emission, and coalescence, followed by more
accretion, bar and lump formation, wave emission, and coalescence.  Bonnell
and Pringle
speculate that hydrodynamics, not wave emission, will drive this
evolution, but that the total energy going into gravitational waves might be
as large as $\sim 10^{-3}M_\odot$.  This corresponds to $h_c \sim 10^{-21}
(10{\rm Mpc}/r)$.

\subsection{Spinning Neutron Stars; Pulsars}
\label{pulsars}

As the neutron star settles down into its final state, its crust begins
to solidify (crystalize). The solid
crust will assume nearly the oblate axisymmetric shape that
centrifugal forces are trying to maintain,
with poloidal
ellipticity $\epsilon_p \propto$(angular velocity of rotation)$^2$.
However, the principal axis
of the star's moment of inertia tensor may deviate from its spin axis
by some small ``wobble angle'' $\theta_w$, and the star may
deviate slightly from axisymmetry about its principal axis; i.e., it may
have a slight ellipticity $\epsilon_e \ll \epsilon_p$ in its equatorial plane.

As this slightly imperfect crust spins, it will radiate gravitational
waves \cite{zimmermann}: $\epsilon_e$ radiates at twice the
rotation frequency, $f=2f_{\rm
rot}$ with
$h\propto \epsilon_e$, and the wobble angle couples to $\epsilon_p$ to
produce waves at $f=f_{\rm rot} + f_{\rm prec}$
(the precessional sideband of the rotation frequency) with amplitude
$h\propto \theta_w \epsilon_p$.  For typical neutron-star masses and
moments of inertia, the wave amplitudes are
\begin{equation}
h \sim 6\times 10^{-25} \left({f_{\rm rot}\over 500{\rm Hz}}\right)^2
\left({1{\rm kpc}\over r}\right)\left({\epsilon_e \hbox{ or }\theta_w\epsilon_p
\over 10^{-6}}\right)\;.
\label{hpulsar}
\end{equation}

The neutron star gradually spins down, due in part to gravitational-wave
emission but perhaps more strongly due to electromagnetic torques associated
with its spinning magnetic field and pulsar emission.
This spin-down reduces the strength of centrifugal forces, and thereby
causes the star's poloidal ellipticity $\epsilon_p$ to decrease, with
an accompanying breakage and resolidification of its crust's crystal structure
(a ``starquake'') \cite{starquake}.
In each starquake, $\theta_w$, $\epsilon_e$, and
$\epsilon_p$ will all change suddenly, thereby changing the amplitudes and
frequencies of the
star's two gravitational ``spectral lines'' $f=2f_{\rm rot}$ and
$f=f_{\rm rot} + f_{\rm prec}$.  After each quake, there should be a
healing period in which the star's fluid core and solid crust, now rotating
at different speeds, gradually regain synchronism.
By monitoring the
amplitudes, frequencies, and phases of the two gravitational-wave
spectral lines, and by
comparing with timing of
the electromagnetic pulsar emission, one might learn much about the
physics of the neutron-star interior.

How large will the quantities $\epsilon_e$ and $\theta_w \epsilon_p$ be?
Rough estimates of the crustal shear moduli and breaking strengths suggest an
upper limit in the range $\epsilon_{\rm max} \sim 10^{-4}$
to $10^{-6}$, and it might be that typical values are
far below this.  We are extremely ignorant, and
correspondingly there is much to be learned from searches for
gravitational waves from spinning neutron stars.

One can estimate the sensitivity of LIGO/VIRGO (or any other broad-band
detector)
to the periodic waves from such a source by multiplying the waves'
amplitude $h$ by the square root of the number of cycles over which one
might integrate to find the signal, $n= f \hat \tau$ where $\hat\tau$ is the
integration time.  The resulting
effective signal strength, $h\sqrt{n}$, is larger than $h$ by
\begin{equation}
\sqrt n = \sqrt{f\hat\tau} = 10^5 \left( {f\over1000{\rm Hz}}\right)^{1/2}
\left({\hat\tau\over4{\rm months}}\right)^{1/2}\;.
\label{ftau}
\end{equation}
This $h\sqrt n$  should be compared (i) to the
detector's rms broad-band noise level for sources in a random direction,
$\sqrt5 h_{\rm rms}$, to deduce a
signal-to-noise ratio, or (ii) to $h_{\rm SB}$ to deduce a
sensitivity for
high-confidence detection when one does not know the waves' frequency in
advance \cite{300yrs}.
Such a comparison suggests that the first interferometers in
LIGO/VIRGO might possibly see waves from nearby spinning
neutron stars, but the odds of success are very unclear.

The deepest searches for these nearly periodic waves will be
performed by narrow-band detectors, whose sensitivities are enhanced
near some chosen frequency at the price of sensitivity loss
elsewhere---e.g., dual-recycled interferometers \cite{meers} or resonant-mass
antennas (Section \ref{resonantbars}).
With ``advanced-detector technology,'' dual-recycled interferometers
might be able to detect with confidence all spinning neutron stars
that have \cite{300yrs}
\begin{equation}
(\epsilon_e \hbox{ or } \theta_w\epsilon_p ) \agt 3\times10^{-10} \left(
{500 {\rm Hz}\over f_{\rm rot}}\right)^2 \left({r\over 1000{\rm pc}}\right)^2.
\label{advancedpulsar}
\end{equation}
There may well be a large number of such neutron stars in our galaxy; but
it is also conceivable that there are none.  We are extremely
ignorant.

Some cause for optimism arises from several physical mechanisms that
might generate radiating ellipticities large compared to
$3\times10^{-10}$:
\begin{itemize}

\item It may be that, inside the superconducting cores of
many neutron stars, there are trapped magnetic fields with mean
strength $B_{\rm core}\sim10^{13}$G or even
$10^{\rm 15}$G.
Because such a field is actually concentrated in flux
tubes with $B = B_{\rm crit} \sim 6\times 10^{14}$G surrounded by
field-free superconductor, its mean pressure is $p_B = B_{\rm core} B_{\rm
crit}/8\pi$.  This pressure could produce a radiating
ellipticity
$\epsilon_{\rm e} \sim \theta_w\epsilon_p \sim p_B/p \sim 10^{-8}B_{\rm
core}/10^{13}$G (where $p$ is the core's material pressure).

\item Accretion onto a spinning neutron star can drive precession (keeping
$\theta_w$ substantially nonzero), and thereby might produce measurably strong
waves \cite{schutz95}.

\item If a neutron star is born rotating very rapidly,
then it may experience a
gravitational-radiation-reaction-driven instability.  In this
``CFS'' (Chandrasekhar, \cite{cfs_chandra} Friedman, Schutz
\cite{cfs_friedman_schutz}) instability,
density waves propagate around the
star in the opposite direction to its rotation, but are dragged forward
by the rotation.  These density waves produce gravitational waves that
carry positive energy as seen by observers far from the star, but
negative energy from the star's viewpoint; and because the
star thinks it is losing negative energy, its density waves get
amplified.  This intriguing mechanism is similar to that by which
spiral density waves are produced in galaxies.  Although the CFS
instability was once thought ubiquitous for spinning stars
\cite{cfs_friedman_schutz,wagoner}, we now
know that neutron-star viscosity will kill it, stabilizing the star and
turning off the waves, when the star's temperature is above some
limit $\sim 10^{10}{\rm K}$ \cite{cfs_lindblom}
and below some limit $\sim 10^9 {\rm K}$
\cite{cfs_mendell_lindblom}; and correspondingly, the instability
should operate only during the first few years of a neutron
star's life, when $10^9 {\rm K} \alt T \alt 10^{10}\rm K$.

\end{itemize}

\subsection{Stochastic Background}
\label{hfbackground}

There should be a stochastic background of gravitational waves in the
high-frequency band produced by processes in the early universe.
Because this background will extend over all gravitational-wave
frequencies, not just high frequencies, we shall delay discussing it
until Section \ref{stochastic}.

\section{LISA: The Laser Interferometer Space Antenna}
\label{lisa}

Turn, now, from the high-frequency band, 1---$10^4$ Hz,
to the low-frequency band, $10^{-4}$---1 Hz.  At present, the most sensitive
gravitational-wave searches at low frequencies are those carried out by
researchers at NASA's Jet Propulsion Laboratory, using microwave-frequency
Doppler tracking of interplanetary spacecraft. These
searches are done at rather low cost, piggy-back on missions
designed for other purposes.  Although they have a possibility
of success, the odds are against them.  Their best past
sensitivities to bursts, for
example, have been $h_{\rm SB} \sim 10^{-14}$, and prospects are good
for reaching $\sim 10^{-15}$---$10^{-16}$ in the next 5 to 10 years.  However,
the strongest low-frequency bursts
arriving several times per year might be no larger than
$\sim 10^{-18}$; and the domain of an assured plethora of signals is
$h_{\rm SB} \sim 10^{-19}$---$10^{-20}$.

To reach into this assured-detection domain will almost certainly
require switching from microwave-frequency tracking of spacecraft (with its
large noise due to fluctuating dispersion in the troposphere and interplanetary
plasma) to optical tracking.  Such a switch is planned for
the 2014 time frame or sooner, when
the European Space Agency (ESA) and/or NASA is likely to fly the {\it
Laser Interferometer Space Antenna}, LISA.

\subsection{Mission Status}
\label{lisastatus}

LISA is largely an outgrowth of 15 years of studies by Peter
Bender and colleagues at the University of Colorado.
In 1990, NASA's
Ad Hoc Committee on Gravitation Physics and Astronomy selected
a LISA-type gravitational-wave detector as its highest priority
in the large space mission category \cite{shapirocommittee}; and since then
enthusiasm for LISA has continued to grow within the
American gravitation community.
Unfortunately, the prospects for NASA to fly such a mission did not
look good in the early 1990s.  By contrast, prospects in Europe looked
much better, so a largely European consortium was put
together in 1993, under the leadership of Karsten Danzmann (Hannover) and
James Hough (Glasgow), to propose LISA to the European Space Agency.  This
proposal has met with considerable success; LISA might well achieve
approval to fly as an ESA Cornerstone Mission around 2014
\cite{cornerstone}.
Members of the American gravitation community and members of the LISA
team hope that NASA will
join together with ESA in this endeavor, and that working jointly, ESA
and NASA will be able to fly LISA considerably sooner than 2014.

\begin{figure}
\vskip11pc
\special{hscale=48 vscale=48 hoffset=14 voffset=7
psfile=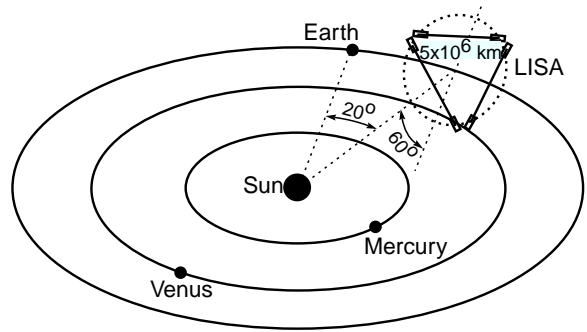}
\caption{LISA's orbital configuration, with LISA magnified in arm length
by a factor $\sim 10$ relative to the solar system.
}
\label{fig:lisa_orbit}
\end{figure}

\subsection{\it Mission Configuration}
\label{lisaconfiguration}

As presently conceived, LISA will consist of six compact, drag-free
spacecraft (i.e. spacecraft that are shielded from buffeting by solar
wind and radiation pressure, and that thus move very nearly on geodesics of
spacetime).  All six spacecraft would be launched simultaneously by a
single Ariane rocket. They
would be placed into the same heliocentric orbit as the Earth
occupies, but would follow 20$^{\rm o}$ behind the Earth; cf.\ Figure
\ref{fig:lisa_orbit}.  The spacecraft would fly in pairs, with each pair
at the vertex of an equilateral triangle that is inclined at an angle of
60$^{\rm o}$ to the Earth's orbital plane. The triangle's arm length would be 5
million km ($10^6$ times larger than LIGO's arms!).  The six spacecraft would
track each other optically, using one-Watt YAG laser beams.  Because of
diffraction
losses over the $5\times10^6$km arm length, it is not feasible to
reflect the beams back and forth between mirrors as is done with LIGO.
Instead, each spacecraft will have its own laser; and the lasers will be
phase locked to each other, thereby achieving the same kind of
phase-coherent out-and-back light travel as LIGO achieves with mirrors.
The six-laser, six-spacecraft configuration thereby functions as three,
partially independent but partially redundant,
gravitational-wave interferometers.

\medskip
\subsection{\it Noise and Sensitivity}
\label{lisanoise}

\begin{figure}
\vskip18.0pc
\special{hscale=46 vscale=46 hoffset=0 voffset=-320
psfile=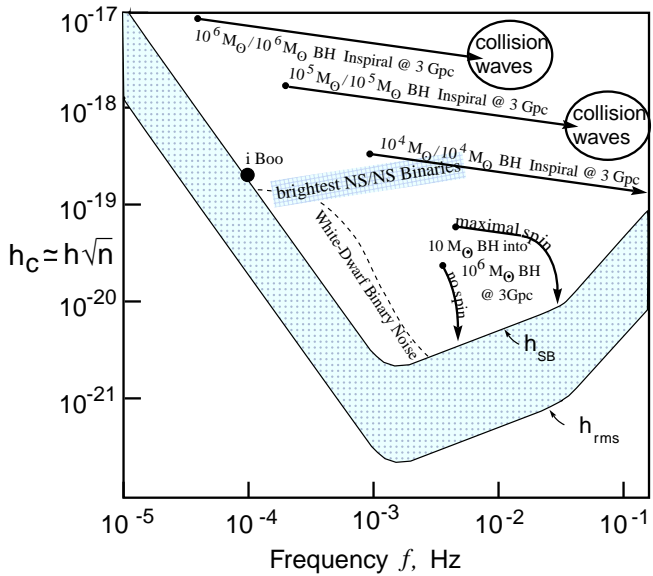}
\caption{LISA's projected broad-band noise $h_{\rm rms}$ and sensitivity
to bursts $h_{\rm SB}$, compared with the strengths of the waves from
several low-frequency sources. [{\it Note:} When members of the LISA team
plot curves analogous to this, they show the sensitivity curve (top of
stippled region) in units of the amplitude of a periodic signal
that can be detected with $S/N=5$ in one year of integration; that
sensitivity to periodic sources is related
to the $h_{\rm SB}$ used here by $h_{\rm SP} = h_{\rm SB}/\protect\sqrt{f\cdot
3\times 10^7\rm{sec}}$.]
}
\label{fig:lisa_noise}
\end{figure}

Figure \ref{fig:lisa_noise} depicts the expected noise and sensitivity of
LISA in the same language as we have used for LIGO (Fig.\
\ref{fig:noisesources}).  The curve at the bottom of the stippled region
is $h_{\rm rms}$, the rms noise, in a bandwidth equal to frequency,
for waves with optimum direction and polarization.  The top of the
stippled region is $h_{\rm SB} = 5\sqrt5 h_{\rm rms}$, the sensitivity
for high-confidence detection ($S/N=5$) of a broad-band burst coming from
a random direction, assuming Gaussian noise.

At frequencies $f\agt 10^{-3}$Hz, LISA's noise is due to photon counting
statistics (shot noise).  The noise curve steepens at $f\sim
3\times10^{-2}$Hz because at larger $f$ than that, the waves' period is
shorter than the round-trip light travel time in one of LISA's arms.
Below $10^{-3}$Hz, the noise is due to buffeting-induced random motions
of the spacecraft
that are not being properly removed by the drag-compensation system.
Notice that, in terms of dimensionless amplitude, LISA's sensitivity is
roughly the same as that of LIGO's first interferometers (Fig.\
\ref{fig:CBStrengthSensitivity}), but at 100,000 times lower frequency.  Since
the waves' energy flux scales as $f^2 h^2$, this corresponds to $10^{10}$
better energy sensitivity than LIGO.

\medskip
\subsection{Observational Strategy}
\label{lisastrategy}

LISA can detect and study, simultaneously, a wide variety of different
sources scattered over all directions on the sky.  The key to
distinguishing the different sources is the different time evolution of
their waveforms.  The key to determining each source's direction, and
confirming that it is real and not just noise, is the manner in which
its waves' amplitude and frequency are modulated by LISA's complicated
orbital motion---a motion in which the interferometer triangle rotates around
its center once per year, and the interferometer plane precesses
around the normal to the Earth's orbit once per year.  Most sources will
be observed for a year or longer, thereby making full use of these
modulations.

\section{Low-Frequency Gravitational-Wave Sources}
\label{lfsources}

\subsection{Waves from Binary Stars}
\label{binaries}

LISA has a large class of guaranteed sources: short-period binary stars
in our
own galaxy.  A specific example is the classic binary 44 i Boo
(HD133640), a $1.35
M_\odot$/$0.68 M_\odot$ system just 12 parsecs from Earth, whose wave
frequency $f$ and characteristic amplitude $h_c = h\sqrt n$
are depicted in Figure \ref{fig:lisa_noise}.  (Here
$h$ is the waves' actual amplitude and $n = f\hat\tau$ is the number of
wave cycles during $\hat \tau =$1 year of signal integration).
Since 44 i Boo lies right on the $h_{\rm SB}$ curve, its signal to
noise ratio in one year of integration should be $S/N = 5$.

To have an especially short period, a binary must be made of especially
compact bodies---white dwarfs (WD), neutron stars (NS), and/or black
holes (BH).  WD/WD binaries are thought to be so numerous that they
may produce a stochastic background of gravitational waves, at the level
shown in Figure \ref{fig:lisa_noise},
that will hide some other interesting waves
from view \cite{wdbinaries}.  Since WD/WD binaries are very dim optically,
their actual numbers are not known for sure; Figure \ref{fig:lisa_noise}
might be an overestimate.

Assuming a NS/NS coalescence rate of 1 each $10^5$ years in our galaxy
\cite{phinney,narayan}, the shortest period NS/NS binary should have a
remaining life of about $5\times 10^4$ years, corresponding to a
gravitational-wave frequency today of $f\simeq 5\times10^{-3}$Hz, an
amplitude (at about $10$kpc distance) $h \simeq 4 \times 10^{-22}$, and a
characteristic amplitude (with one year of integration time) $h_c \simeq
2\times 10^{-19}$.  This is depicted in Figure \ref{fig:lisa_noise} at the
right edge of the region marked ``brightest NS/NS binaries''. These
brightest NS/NS binaries can be studied by LISA with
the impressive signal to noise ratios $S/N \sim 50$ to $500$.

\subsection{Waves from the Coalescence of Massive Black Holes in
Distant Galaxies}
\label{lfbhcoalescence}

LISA would be a powerful instrument for studying massive black holes in
distant galaxies.  Figure \ref{fig:lisa_noise} shows, as examples, the
waves from several massive black hole binaries at 3Gpc distance from Earth (a
cosmological redshift of unity).  The waves sweep upward in frequency
(rightward in the diagram) as the holes spiral together.  The black dots
show the waves' frequency one year before the holes' final collision and
coalescence, and the arrowed lines show the sweep of frequency and
characteristic amplitude $h_c = h\sqrt n$ during that last year.  For
simplicity, the figure is restricted to binaries with equal-mass black
holes:
$10^4M_\odot / 10^4 M_\odot$, $10^5M_\odot / 10^5 M_\odot$,
and $10^6M_\odot / 10^6 M_\odot$.

By extrapolation from these three examples, we see that LISA can
study much of the last year of inspiral, and the waves
from the final collision and coalescence,
whenever the holes' masses are in the range $3\times 10^4 M_\odot
\alt M \alt 10^8
M_\odot$.  Moreover, LISA can study the final coalescences with remarkable
signal to noise ratios: $S/N \agt 1000$.
Since these are much larger $S/N$'s than LIGO/VIRGO is likely to achieve,
we can expect LISA to refine the experimental understanding of black-hole
physics, and of highly nonlinear vibrations of warped spacetime,
which LIGO/VIRGO initiates---{\it provided} the rate of massive
black-hole coalescences is of order one per
year in the Universe or higher.  The rate might well be that high, but
it also might be much lower.

By extrapolating Figure \ref{fig:lisa_noise} to lower BH/BH masses, we
see that LISA can observe the last few years of inspiral, but not the
final collisions, of binary black holes in the range
$100M_\odot \alt M \alt 10^4 M_\odot$, out to cosmological distances.

Extrapolating the BH/BH curves to lower frequencies using the
formula (time to final coalescence$)\propto f^{-8/3}$, we see that
equal-mass BH/BH binaries enter LISA's frequency band roughly 1000 years
before their final coalescences, more or less independently of their
masses, for the range $100 M_\odot \alt M \alt 10^6 M_\odot$.  Thus, if the
coalescence rate were to turn out to be one per year, LISA would see
roughly 1000 additional massive binaries that are slowly spiraling
inward, with inspiral rates $df/dt$ readily measurable.  From the inspiral
rates, the amplitudes of the two polarizations, and the waves' harmonic
content, LISA can determine each such binary's luminosity distance,
redshifted chirp mass $(1+z)M_c$, orbital inclination,
and eccentricity; and from the waves' modulation by LISA's orbital
motion, LISA can learn the direction to the binary with an accuracy of
order one degree.

\subsection{Waves from Compact Bodies Spiraling into Massive Black
Holes in Distant Galaxies}
\label{lfbhinspiral}

When a compact body with mass $\mu$ spirals into a much more massive black
hole with mass $M$, the body's orbital energy $E$ at fixed frequency
$f$ (and correspondingly at fixed orbital radius $a$)
scales as $E \propto \mu$,
the gravitational-wave luminosity $\dot E$ scales as
$\dot E \propto \mu^2$, and the time to
final coalescence thus scales as $t \sim E/\dot E \propto 1/\mu$.  This
means that the smaller is $\mu/M$,
the more orbits are spent in the hole's strong-gravity region, $a\alt
10GM/c^2$, and thus the more detailed and accurate will be the map of the
hole's spacetime geometry, which is encoded in the emitted waves.

For holes observed by LIGO/VIRGO, the most extreme mass ratio that we
can hope for is $\mu/M \sim 1M_\odot/300 M_\odot$, since for $M>300M_\odot$ the
inspiral waves are pushed to frequencies below the LIGO/VIRGO band.
This limit on $\mu/M$ seriously constrains the accuracy with which
LIGO/VIRGO can hope to map out the spacetime geometries of black
holes and test the black-hole no-hair theorem \cite{ryan_finn_thorne}
(end of Section \ref{cbwaveforms}).
By contrast, LISA can observe the final inspiral waves from objects of
any mass $M\agt 0.5M_\odot$ spiraling into holes of mass $3\times 10^5 M_\odot
\alt M \alt 3\times10^7M_\odot$.

Figure \ref{fig:lisa_noise} shows the
example of a $10M_\odot$ black hole spiraling into a $10^6M_\odot$ hole
at 3Gpc distance.  The inspiral orbit and waves are strongly influenced
by the hole's spin.  Two cases are shown \cite{finn_thorne}:
an inspiraling circular orbit
around a non-spinning hole, and a prograde, circular, equatorial orbit
around a maximally spinning hole.
In each case the dot at the upper left end of the
arrowed curve is the frequency and characteristic amplitude one year
before the final coalescence.  In the nonspinning case, the small hole
spends its last year spiraling inward from $r\simeq 7.4 GM/c^2$
(3.7 Schwarzschild
radii) to its last stable circular orbit at $r=6GM/c^2$ (3 Schwarzschild
radii).  In the maximal spin case, the last year is spent traveling from
$r=6GM/c^2$ (3 Schwarzschild radii) to the last stable orbit at $r=GM/c^2$
(half a
Schwarzschild radius).  The $\sim 10^5$ cycles of waves during this last
year should carry, encoded in themselves, rather accurate values for
the massive hole's lowest few multipole moments \cite{ryan}.  If the
measured moments satisfy the ``no-hair'' theorem (i.e., if they are all
determined uniquely by the measured mass and spin in the manner of the
Kerr metric), then we can be sure the central body is a black hole.  If
they violate the no-hair theorem, then (assuming general relativity is
correct), either the central body was not a black hole, or an accretion
disk or other material was perturbing its orbit \cite{chakrabarti}.
{}From the evolution of the waves one can hope to determine which is
the case, and to explore the properties of the central body and its
environment \cite{ryan_finn_thorne}.

Models of galactic nuclei, where massive holes reside, suggest that
inspiraling stars and small holes typically will be in rather eccentric
orbits \cite{hils_bender}.  This is because they get injected into such
orbits via
gravitational deflections off other stars, and by the time gravitational
radiation reaction becomes the dominant orbital driving force, there is
not enough inspiral left to fully circularize their orbits.  Such orbital
eccentricity will complicate the waveforms and complicate the extraction
of information from them.  Efforts to understand the emitted waveforms
are just now getting underway.

The event rates for inspiral into massive black holes are not at all
well understood.  However, since a significant fraction of all galactic
nuclei are thought to contain massive holes, and since white dwarfs and
neutron stars, as well as small black holes, can withstand tidal
disruption as they plunge toward the massive hole's horizon, and since
LISA can see inspiraling bodies as small as $\sim 0.5 M_\odot$ out to
3Gpc distance, the event rate is likely to be interestingly large.

\section{The Stochastic Gravitational-Wave Background}
\label{stochastic}

Processes in the early universe should have produced a stochastic
background of gravitational waves that extends through the entire
frequency range from extremely low frequencies $f\sim 10^{-18}$ Hz to
the high-frequency band $f\sim 1$---$10^4$Hz and beyond.

\subsection{Primordial Gravitational Waves}

The most interesting background would be that produced in the big bang itself.
Zel'dovich and Novikov have estimated \cite{zeldovich} that
the optical thickness
of primordial matter to gravitational waves has been small compared to
unity at all times since the Planck era, when space and time came into
being, and that
therefore primordial gravitational waves (by contrast with
electromagnetic) should not have been thermalized by interactions
with matter.
On the other hand (as Grishchuk has shown \cite{grishchuk}),
whatever might have been the state of the graviton field when it emerged
from the big bang's Planck era, it should have interacted with the subsequent,
early-time expansion of the universe to produce, via parametric
amplification, a rich spectrum of stochastic waves today.
The details of that spectrum depend on what emerged from the Planck era
and on the evolution $a(t)$ of the universal expansion factor at early
times.

The gravitational-wave spectrum is generally described by the quantity
$\Omega_g (f) =($energy density in a bandwidth equal to
frequency $f$)/(energy density required to close the universe); cf.\
Section \ref{vlf}.  The
observed quadrupolar anisotropy of the cosmic microwave radiation places
a limit $\Omega_g \alt 10^{-9}$ at $f\sim 10^{-18}$Hz (Section
\ref{elf}). It is
fashionable to extrapolate this limit to higher frequencies
by {\it assuming} that the
graviton field emerged from the Planck era in its vacuum state, and {\it
assuming} that the
universal expansion $a(t)$ was that of an inflationary era $a\propto
e^{Ht}$ for some constant $H$, followed by a radiation-dominated Friedman
era $a\propto t^{1/2}$, followed by the present matter-dominated
era $a\propto t^{2/3}$.  This standard model produces a flat
spectrum  $\Omega_g$ independent of $f$ for all waves that
entered our cosmological horizon during the radiation-dominated era,
which means
at all frequencies from $\sim 10^{-16}$Hz up
through the high-frequency band and somewhat beyond. The observational
limit at $10^{-18}$Hz implies that this constant value is $\Omega_g
\alt 3\times 10^{-14}$ \cite{krauss_white,primordial_spectrum}.  So weak a
background cannot be detected by LIGO in the high-frequency band, nor by
LISA at low frequencies, nor by pulsar timing at very low frequencies.
LIGO's limiting sensitivities will correspond to
$\Omega_g \sim \hbox{(a few)}\times 10^{-7}$ at $f\sim 10^2$ Hz
for the first
interferometers, and $\Omega_g \sim \hbox{(a few)}\times 10^{-10}$ for
advanced interferometers \cite{flanagan}; LISA's sensitivity
will correspond to $\Omega_g \sim \hbox{(a few)}\times 10^{-10}$ at
$f\sim 10^{-3}$ Hz; and the present pulsar timing measurements
correspond to $\Omega_g \sim \hbox{(a few)}\times 10^{-8}$
at $f \sim 4\times 10^{-9}$Hz (Section
\ref{vlf} and Ref.\ \cite{kaspi_taylor}).

On the other hand, if the
graviton field did not begin in its vacuum state, or if the equation of
state in the very early Friedman era was stiffer
than that of radiation, then the primordial backgrounds at high, low, and
very low frequencies could be significantly
stronger than $\Omega \sim 3\times 10^{-14}$, and could be strong
enough to detect.

\subsection{Waves from Phase Transitions in the Early Universe}

A stochastic background could also have been produced by phase
transitions in the early universe \cite{witten,turner}. No known phase
transition would put its waves into the high-frequency band, and even
hypothetical phase transitions, optimized at high-frequencies, can be only
strong enough for marginal detection by advanced LIGO
interferometers.  The prospects for LISA are a little better: A strongly
first-order electroweak phase transition could produce low-frequency
waves strong
enough for LISA to detect \cite{turner}.

\subsection{Waves from Cosmic Strings}

If cosmic strings \cite{vilenkin}
were produced in the early universe in as large
numbers as some theorists have suggested
\cite{zeldovich_strings,vilenkin_strings}, their vibrations would produce
a gravitational wave spectrum that is frequency
independent, $\Omega_g = \hbox{const}$, from below the very low-frequency
band where pulsar timing operates, through LISA's low-frequency band,
and on into and through the high-frequency band.
Theory suggests \cite{vachaspati} that such waves could be as strong as
$\Omega_g
\sim 10^{-7}$---a level that is already being constrained by pulsar
timing observations (Section \ref{vlf}).  LIGO's first interferometers
will operate at this
same level, and by the time LIGO's advanced interferometers and LISA reach
$\Omega_g \sim \hbox {(a few)}\times10^{-10}$, pulsar timing
might be in that same ballpark.

\medskip

To summarize: There are known mechanisms that could easily produce a
measurable stochastic background in the high, low, and very-low
frequency bands. However, the odds of the background being that large,
based on currently
fashionable ideas, are not great.  Despite this, a vigorous effort to
detect background waves and to map their spectrum will surely be made, since
the cosmological implications of their discovery could be profound.

\section{Conclusion}
\label{conclusion}

It is now 35 years since Joseph Weber initiated his pioneering
development of gravitational-wave detectors \cite{weber} and 25 years
since Forward \cite{forward} and Weiss \cite{weiss}
initiated work on interferometric detectors.
Since then, hundreds of
talented experimental physicists have struggled to improve the
sensitivities of these instruments.  At last, success is in sight.  If
the source estimates described in this review article are approximately
correct, then the planned interferometers should detect the first waves
in 2001 or several years thereafter, thereby opening up this rich new
window onto the Universe.

\section{Acknowledgments}
\label{acknowledgments}

For insights into the rates of coalescence of compact binaries, I thank
Sterl Phinney.  My group's research on gravitational waves
and their relevance to LIGO/VIRGO and LISA is supported in part
by NSF grants AST-9417371 and PHY-9424337 and by NASA grant NAGW-4268.
Portions of this review article were adapted from my Ref.\
\cite{nishinomiya}.

\end{document}